\documentclass[fleqn,usenatbib]{mnras}
\usepackage{amsmath,amssymb}
\usepackage{xymtex}
\usepackage{newtxtext,newtxmath}
\usepackage[T1]{fontenc}
\usepackage{graphicx}	
\usepackage{amsmath}	
\usepackage{amssymb}	
\usepackage{color}
\usepackage{float}
\usepackage{placeins}
\usepackage{multicol}
\usepackage{comment}
\usepackage{soul}
\usepackage{threeparttable}

\def\be{\begin{equation}} 
\def\ee{\end{equation}}

\def\msun{{\Msun}}

\def\HI{\hbox{H~$\scriptstyle\rm I\ $}}

\def\gsim{\lower.5ex\hbox{\gtsima}} 
\def\lsim{\lower.5ex\hbox{\ltsima}} \def\gtsima{$\; \buildrel > \over 
\sim \;$} \def\ltsima{$\; \buildrel < \over \sim \;$} \def\prosima{$\; 
\buildrel \propto \over \sim \;$} \def\gsim{\lower.5ex\hbox{\gtsima}} 
\def\lsim{\lower.5ex\hbox{\ltsima}} 
\def\simgt{\lower.5ex\hbox{\gtsima}} 
\def\simlt{\lower.5ex\hbox{\ltsima}} 
\def\simpr{\lower.5ex\hbox{\prosima}} \def\la{\lsim} \def\ga{\gsim} 
  
 \def\gtsima{$\; \buildrel > \over \sim \;$} 
\def\ltsima{$\; \buildrel < \over \sim \;$} 
\def\gsim{\lower.5ex\hbox{\gtsima}} 
\def\lsim{\lower.5ex\hbox{\ltsima}} 
\def\simgt{\lower.5ex\hbox{\gtsima}} 
\def\simlt{\lower.5ex\hbox{\ltsima}} 
\def\simpr{\lower.5ex\hbox{\prosima}} 
\def\la{\lsim} 
\def\ga{\gsim}

\def\msun{\,{\rm \Msun}}

\def\E3{{\cal E}_{\rm g}^{III}}

\def\Msun{\rm M_\odot}

\def\zsun{\rm Z_\odot}

\def\Msun{\rm M_\odot}

\def\M*{M_*}

\def\Z*{Z_*}
\def\L*{L_*}

\def\fej{f_*^{\rm ej}}
\def\feff{f_*^{\rm eff}}

\def\fesc{f_\mathrm{esc}}

\newcommand{\quotes}[1]{``#1''}
\newcommand\textlcsc[1]{\textsc{\MakeLowercase{#1}}}

\title[Metallicity scaling relations during the EoR]{Astraeus V: The emergence and evolution of metallicity scaling relations during the Epoch of Reionization} 
\author[G. Ucci et al.]{Graziano Ucci$^1$\thanks{graziano.ucci@sns.it}, Pratika Dayal$^1$\thanks{p.dayal@rug.nl}, Anne Hutter$^1$, Chiaki Kobayashi$^2$, Stefan Gottl\"ober$^3$,
\newauthor Gustavo Yepes$^{4,5}$, Leslie Hunt$^6$, Laurent Legrand$^1$ and Crescenzo Tortora$^{6,7}$   \\
$^{1}$Kapteyn Astronomical Institute, University of Groningen, P.O. Box 800, 9700 AV Groningen, The Netherlands\\
$^{2}$Centre for Astrophysics Research, Department of Physics, Astronomy and Mathematics, University of Hertfordshire, College Lane,\\
Hatfield, Hertfordshire AL10 9AB, UK\\
$^{3}$Leibniz-Institut f\"ur Astrophysik, An der Sternwarte 16, 14482 Potsdam, Germany\\
$^{4}$Departamento de Fısica Teorica, Modulo 8, Facultad de Ciencias, Universidad Autonoma de Madrid, 28049 Madrid, Spain\\
$^{5}$CIAFF, Facultad de Ciencias, Universidad Autonoma de Madrid, 28049 Madrid, Spain\\
$^{6}$INAF -Osservatorio Astrofisico di Arcetri, Largo E. Fermi 5, 50125 - Firenze, Italy\\
$^{7}$INAF -Osservatorio Astronomico di Capodimonte, Salita
Moiariello 16, 80131 - Napoli, Italy}

\date{Accepted XXX. Received YYY; in original form ZZZ}

\pubyear{2021}

\begin{document}
\label{firstpage}
\pagerange{\pageref{firstpage}--\pageref{lastpage}}
\maketitle

\begin{abstract}
In this work, we have implemented a detailed physical model of galaxy chemical enrichment into the \textlcsc{Astraeus} (semi-numerical rAdiative tranSfer coupling of galaxy formaTion and Reionization in N-body dark matter simUlationS) framework which couples galaxy formation and reionization in the first billion years. Simulating galaxies spanning over 2.5 orders of magnitude in halo mass with $M_h \sim 10^{8.9-11.5}\Msun$ ($M_h \sim 10^{8.9-12.8}\Msun$) at $z \sim 10 ~ (5)$, we find: (i) smooth-accretion of metal-poor gas from the intergalactic medium (IGM) plays a key role in diluting the interstellar medium (ISM) metallicity which is effectively restored due to self-enrichment from star formation; (ii) a redshift averaged gas-mass loading factor that depends on the stellar mass as $\eta_g \approx 1.38 ({M_*}/{10^{10}\msun})^{-0.43}$; (iii) the mass-metallicity relation is already in place at $z \sim 10$ and shows effectively no redshift evolution down to $z \sim 5$; (iv) for a given stellar mass, the metallicity decreases with an increase in the star formation rate (SFR); (v) the key properties of the gas-phase metallicity (in units of 12+log(O/H), stellar mass, SFR and redshift are linked through a high-redshift fundamental plane of metallicity (HFPZ) for which we provide a functional form; (vi) the mass-metallicity-SFR relations are effectively independent of the reionization radiative feedback model for $M_* \gsim 10^{6.5}\Msun$ galaxies; (vii) while low-mass galaxies ($M_h \lsim 10^9\Msun$) are the key contributors to the metal budget of the IGM at early times, higher mass halos provide about 50\% of the metal budget at lower-redshifts. 
\end{abstract}

\begin{keywords}
galaxies: high-redshift, evolution, formation, abundances - dark ages, reionization - methods: numerical
\end{keywords}

\section{Introduction}
The chemical evolution of the interstellar medium (ISM) is a fundamental indicator of the key baryonic processes that govern galaxy formation and evolution. In addition to the entire star formation history (SFH), the metal content of a galaxy crucially depends on a number of (environmental-dependent) quantities including: the metals and gas gained by mergers and through inflows from the circumgalactic (CGM) and intergalactic medium (IGM) and the metals and gas lost via supernova (SN) winds that depend on the instantaneous star formation rate (SFR), through black hole-powered outflows as well as being swept up into the next generation of star formation (astration). Therefore, connecting the metallicity (ratio of metal mass-to-gas mass) with the underlying galaxy properties (such as stellar mass and SFR) can provide useful insights into galaxy growth throughout cosmic history \citep{maiolino2019}. 

The metallicity usually denotes the ISM gas-phase metallicity and is typically expressed through the gas-phase oxygen abundance (the most abundant heavy element) as 12+log(O/H). The relation between the gas-phase oxygen abundance and the stellar mass is referred to as the Mass-Metallicity relation (MZR). Extensive observations have been undertaken to establish this relation in the local Universe \citep{tremonti2004,savaglio2005,lee2006,kewley2008,zahid2011,berg2012,andrews2013,ly2015,ly2016,blanc2019,curti2020}, for intermediate redshifts \citep[i.e., $1 \lesssim z \lesssim 3$;][]{wuyts2012,henry2013,kulas2013,cullen2014,maier2014,steidel2014,zahid2014a,zahid2014b,kacprzak2015,sanders2015,kacprzak2016,wuyts2016,sanders2018}, and even for redshifts up to $z \sim 3.5$ \citep{maiolino2008,mannucci2009,mannucci2010,hunt2012,belli2013,troncoso2014,hunt2016,onodera2016,sanders2020}. This has led to an emerging consensus that the MZR evolves up to $z \sim 3.5$ such that at a fixed stellar mass, the metallicity declines with increasing redshift. Despite this progress, the existence and evolution of the MZR at higher redshifts remain open questions. This is because metallicity measurements are almost exclusively based on the rest-frame optical (strong) emission lines which are redshifted out of the spectral range of current spectrographs for $z \gtrsim 5$. Over the next years, forthcoming facilties such as the James Webb Space Telescope (JWST) will be crucial in extending such observed relations well into the epoch of reionization (EoR). Additionally, over the past decade, it has also been shown that the MZR is actually the two-dimensional projection of an underlying three-dimensional relation linking the stellar mass, gas-phase metallicity and the instantaneous SFR. This ``fundamental metallicity relation" (FMR) has now been widely explored, both from the observational \citep[e.g.][]{ellison2008,lara-lopez2010,mannucci2010,hunt2012,salim2014, hunt2016, cresci2019, curti2020} and theoretical perspectives \citep[e.g.][]{yates2012,dayal2013,lilly2013, yates2014, dave2012, hunt2016b}. 

As noted, chemical evolution in galaxies is affected by a great number of processes including accretion of gas from the IGM, star formation inside galaxies fuelled by the their gas reservoir and their associated yields, feedback from internal (e.g. SN feedback) and external sources (reionization), gas, stars and metals merging via major and minor mergers from previous galaxy generations, and ejection of metal-rich gas into the CGM/IGM that could later be re-accreted onto the galaxy, the metal-dust cycle and the presence of accreting black holes. A number of theoretical models have been developed to study different aspects of these effects as well as their impact on the metal enrichment. These range from semi-analytical models \citep[SAMs; e.g.][]{delucia2004, somerville2008, croton2016, mutch2016} to hydrodynamic simulations \citep[e.g.][]{kobayashi2007, tornatore2007, maio2015,taylor2016, okamoto2017, torrey2019, ma2016, derossi2017, langan2020} to zoom-in simulations \citep{katz2019, pallottini2019}. As might be expected, these different approaches have very different strengths: while SAMs can be used to explore a wide range of physical processes for the entire galaxy population, they can not model the ISM in the exquisite detail that is being achieved by zoom-in simulations. The latter, however, can, at most, model a few galaxies given their computational requirements. Finally, while hydrodynamic simulations can be used to model  representative galaxy populations through time, most of these do not fully couple galaxy formation with reionization. This could be particularly relevant for the metallicity of low-mass galaxies at high-$z$ since a growing body of work shows that the ultra-violet background (UVB) created during reionization can (through photo-evaporation) suppress the gas mass and hence the star formation rates in low-mass galaxies with halo mass $M_h\lesssim10^9\msun$ \citep{Gnedin2000, Hasegawa2013, Gnedin2014, Pawlik2015, ocvirk2016, Ocvirk2018, Dawoodbhoy2018, Katz2019b, Wu2019, astraeus1}. As a result, the slope and normalization of the MZR as well as its redshift evolution remain a matter of debate at high-$z$. 

The aim of this work is to study the key physics determining the emergence and redshift evolution of the MZR in the EoR. We use the \textlcsc{Astraeus} framework \citep{astraeus1} that self-consistently couples a state-of-the-art N-body simulation (very small multi-dark; {\sc vsmd}) with a semi-analytic model of galaxy formation \citep[{\sc delphi};][] {dayal2013} and a semi-numerical reionization scheme \citep[{\sc cifog};][]{hutter2018}. This work focuses on a major update of this code through the detailed implementation of the chemical enrichment of galaxies. The key strengths of this work lie in: (i) an exploration of the entire plausible range of radiative feedback models (ranging from a weak, time-delayed to a strong instantaneous reduction of gas in the galaxy) that are fully coupled to the underlying galaxy population; and (ii) the inclusion of the realistic stellar yields. Indeed,  \textlcsc{Astraeus} is currently the only semi-analytic model that includes the latest state-of-the-art yields from \citet{kobayashi2020} so far. This yield set can reproduce the observations not only for oxygen but also for most of all stable elements (up to uranium) self-consistently for the Milky Way. 

The paper is structured as follows. In Section \ref{sec_model} we describe the \textlcsc{Astraeus} framework including a full description of the underlying N-body simulation, the galaxy formation model with complete details of chemical enrichment and the coupling to the different radiative feedback models explored. In Sec. \ref{sec_assembly}, we study the assembly of the gas and metal content of early galaxies before discussing the mass loading of gas and metals in Sec. \ref{sec:loading}. We then explore the relation between the stellar mass, metallicity and the specific star formation rate in Sec. \ref{sec:sfr} before discussing the emergence of the MZR in Sec. \ref{sec:HFPZ} and its redshift evolution in Sec. \ref{sec:zev}. We then discuss the resulting metal enrichment of the IGM in the first billion years in Sec. \ref{sec:igm_met} before ending with our conclusions in Sec. \ref{sec:conc}.

\section{Theoretical model}
\label{sec_model}
As in the preceding papers of this series, in this work we use the \textlcsc{Astraeus} (semi-numerical rAdiative tranSfer coupling of galaxy formaTion and Reionization in N-body dArk mattEr simUlationS) framework that couples a state-of-the-art N-body simulation run as part of the Multi-dark project\footnote{See www.cosmosim.org for further information about the Multi-dark suite of simulations and access to the simulations database.} ({\it Very small multi-dark Planck}; \textlcsc{VSMDPL}) with a slightly modified version of the \textlcsc{Delphi} semi-analytic model of galaxy formation \citep{dayal2014} and the \textlcsc{CIFOG} (Code to compute ionization field from density fields and source catalogue) semi-numerical reionization scheme \citep{hutter2018}. We briefly describe the model here and interested readers are referred to \citet{astraeus1} for complete details. The major update of \textlcsc{Astraeus} in this work is the implementation of a complete physical model to track the chemical evolution of galaxies at $z \gsim 4.5$.

The underlying dark matter-only N-body simulation has been run using the \textlcsc{gadget-2} Tree+PM (particle mesh) N-body code \citep{springel2005, klypin2016}. It has a box side length of $160h^{-1}$ cMpc and follows the trajectories of $3840^3$ particles, with each particle having a dark matter mass of 6.2 $\times10^6h^{-1}\msun$. In total, 150 simulation snapshots have been stored from $z=25$ to $z=0$, with $74$ being saved between $z=25$ and $z=4.5$. For all snapshots, halos and subhalos down to $20$ particles have been identified using the \textlcsc{rockstar} phase-space halo finder \citep{behroozi2013}. We limit ourselves to using halos with at least 100 particles\footnote{The resolution study in Appendix C of \citet{astraeus1} shows that the physical properties of the galaxies in our simulations converge for halos with at least 50 particles.} in this work resulting in a minimum halo mass of $M_h = 10^{8.95}\msun$. Merger trees have been generated from the \textlcsc{rockstar} catalogues by using \textlcsc{consistent trees} \citep{behroozi2013_trees} which have been resorted to local horizontal (sorted on a redshift-by-redshift-basis within a tree) merger trees using the \textlcsc{cutnresort} module within the \textlcsc{astraeus} pipeline \citep{astraeus1}. In addition, for all snapshots, the dark matter density fields have been re-sampled to a $512^3$ grid which are used as input files for the \textlcsc{Astraeus} code.

\textlcsc{Astraeus} includes all the key baryonic processes of gas accretion, gas and stellar mass being brought in by mergers, star formation and Type II supernova (SNII) feedback and the impact of radiative feedback from the patchy UVB generated during reionization. The UVB can (i) photo-evaporate gas from small gravitational potentials and (ii) increase the Jeans mass for galaxy formation (which reduces the amount of gas accreted from the IGM). A combination of these two effects leads to a reduction in the gas content of low-mass halos in ionized regions. At each redshift step, these baryonic processes are coupled to the merger- and accretion-driven growth of the dark matter halos obtained from the N-body simulation as detailed in what follows. 

Throughout this work we use a Salpeter initial mass function \citep[IMF,][]{salpeter1955} between 0.1 and 100 $\Msun$. This yields only slightly lower metallicities as compared to a Kroupa IMF over 0.01-100 $\msun$; e.g. see \citet{kobayashi2010} for the IMF dependence on the  MZR and \citet{vincenzo2016} for metal yields as a function of the IMF. Finally, the cosmological model used in this work corresponds to the $\Lambda$CDM Universe with dark matter, dark energy and baryonic density parameter values of ($\Omega_{\rm m },\Omega_{\Lambda}, \Omega_{\rm b}) = (0.31,0.69,0.048)$, a Hubble constant $H_0 = 100h= 68 \, {\rm km\, s^{-1} Mpc^{-1}}$, a primordial spectral index $n_s=0.96$ and a spectral normalisation $\sigma_8=0.83$, consistent with the latest results from the \textlcsc{Plank} collaboration \citep{planck2016}.

\begin{itemize}
	\item {\it Dark matter accretion and mergers}: in the \textlcsc{Astraeus} framework, a galaxy with a halo mass $M_h(z)$ that is a starting leaf (i.e. has no progenitors), can smoothly accrete a gas mass of $M_g^{\rm acc}(z)$ from the surrounding IGM, corresponding to the cosmological baryon-to-dark matter ratio. On the other hand, halos that have progenitors gain gas both through mergers and accretion. For halos in ionized regions, this gas mass can be reduced due to reionization feedback. The initial gas mass, $M_g^i(z)$, that can be retained inside a halo of mass $M_h$ can then be expressed as:
	\begin{equation}
	M_g^i(z) = \min \left[M_g^{\rm mer}(z)+M_g^{\rm acc}(z),\ f_g \frac{\Omega_b}{\Omega_m}M_h(z) \right],
	\label{eq:mgasini}
	\end{equation}
	where $M_g^{\rm mer}(z)$ is the gas mass brought in by mergers (this has a value 0 for starting leaves) and $f_g$ is the gas fraction that remains available for star formation in the presence of an UVB. The value of $f_g$ crucially depends on a number of parameters including the characteristic mass for radiative feedback ($M_c$; halo mass at which half of the baryonic mass is retained), the host halo mass and its redshift, and the redshift at which the halo was first irradiated by the UVB \citep[see Sec. \ref{sec_yields} of][]{astraeus1}.\\
	
	\item {\it Star formation and SN feedback}: at a given snapshot, the initial gas mass can form stars with an {\it effective} efficiency, $\feff$, such that the newly formed stellar mass is $M_*^{\rm new}(z) = M_{g}^i(z) \feff$. The explosion of high-mass stars, in addition to chemically enriching the ISM, inject thermal and kinetic energy into it. This energy can heat and expel gas from the galaxy. In our model we assume that each SNII produces an energy equal to $E_{51} = 10^{51}{\rm erg}$ of which a fraction ($f_w$) couples to the gas and drives outflows. In this work we implement the \quotes{delayed SN} scheme that accounts for the mass-dependent lifetimes of stars \citep{padovani1993}. The value of $\feff$ is then the minimum between the star formation efficiency that produces enough SNII energy to unbind the rest of the gas from the halo potential ($\fej$) and an upper threshold ($f_* \sim 1-3\%$) such that $\feff = \min[ f_*, \fej]$. This star formation is assumed to be uniformly distributed over the entire redshift step resulting in a SFR: 
    \begin{equation}
    {\rm SFR}(t) = \frac{M_*^{\rm new}(t)}{\Delta t}
    \end{equation}
    where $\Delta t$ is the time interval between two consecutive snapshots at redshifts (i.e. $z + \Delta z$ and $z$).\\

	\item{\it Patchy UVB and reionization feedback}: in order to simulate (patchy) reionization, we use the \textlcsc{Cifog} code \citep{hutter2018}. This is a MPI-parallelised, semi-numerical reionization code that uses the ionizing emissivities and positions of the underlying galaxy population and the local gas density (on a $512^3$ grid) to yield the photoionization rate, residual neutral hydrogen fraction and recombination rate of each cell. For galaxies lying in reionized regions, we calculate the fraction of gas mass they can retain after radiative feedback; galaxies in neutral regions are naturally unaffected by reionization feedback. In this work we consider 2 out of the 3 cases studied in \citet{astraeus2}: the \textit{Photoionization} and \textit{Jeans mass} models as shown in Table \ref{table:models}. These are our \quotes{extreme} cases in terms of the characteristic mass; both assume a constant value of the escape fraction of ionizing photons ($f_{\rm esc}$). The value of $M_c$ is minimum for the \textit{Photoionization} model where we account for gas reacting to an increase in the IGM temperature (to $T_{\rm IGM}=10^4$ K) on a dynamical timescale. $M_c$ has its maximum value in the \textit{Jeans mass} model where the IGM is assumed to be heated to $4\times10^4$~K via photo-heating upon ionization and the rise in temperature is assumed to translate instantaneously into a higher Jeans mass.
\end{itemize}

\begin{table*}
	\caption{For the UV feedback model shown in column 1, we show the value of the threshold star formation efficiency (column 2), the fraction of SNII energy that can couple to gas (column 3), the escape fraction of \HI ionizing photons (column 4), the IGM temperature in ionized regions (column 5) and the characteristic halo mass ($M_c$) at $z = 7$ for which halos can retain half of their gas mass after UV feedback assuming a reionization redshift $z_{\rm reion} = 8~ (12)$ (column 6). }
	\begin{threeparttable}
	\begin{tabular}{c c c c c c}
		\hline\hline
		Model & ${f_*}$ & $f_w$ & $f_{\rm esc}$ & $T_{\rm IGM}$ & log ($M_c / \Msun$)\\
		\hline
		\textit{Photoionization} & 0.01 & 0.2 & 0.215 & $\sim 4 \times 10^4$~K\footnotemark[1] & 9.09\footnotemark[2] (7.86)\\
		\textit{Jeans mass} & 0.01 & 0.2 & 0.285 & $4 \times 10^4$~K & 9.52 (9.52)\\
		\hline
	\end{tabular}
	\begin{tablenotes}
		\item[1]{\scriptsize $T_{\rm IGM}$ is actually given by the \HI photoionization rate $\Gamma_{\rm HI}$ which is about $\sim10^{-12.3}$s$^{-1}$}
		\item[2]{\scriptsize \citet{sobacchi2013a}}
	\end{tablenotes}
	\end{threeparttable}
	\label{table:models}
\end{table*}

Our model has a total of three free parameters: the threshold star formation efficiency ($f_*$), the fraction of SNII energy that couples to gas ($f_w$) and the escape fraction of ionizing photons from the galactic environment into the IGM ($\fesc$). These are tuned to reproduce the key observables for: (i) {\it galaxies} - including the evolving ultra-violet luminosity function (UVLF) and stellar mass function (SMF), and the redshift evolution of the star formation rate density (SFRD) and the stellar mass density (SMD) at $z \gsim 5$; and (ii) {\it reionization} - including the electron scattering optical depth and the ionizing history constraints from quasars, Lyman Alpha Emitters (LAEs) and Gamma Ray Bursts (GRBs). The values of the model free parameters, the IGM temperature in ionized regions and the characteristic halo mass for each of the UV feedback models used in this work are reported in Table \ref{table:models}. Interested readers are refereed to (Figs. 2 and C1) in \citet{astraeus1} to see a comparison of our model results with the observed evolving UVLF and the SMF, respectively.

We now detail the implementation of chemical feedback in Secs. \ref{sec:accretion}-\ref{sec:gasmet} that follow.

\subsection{Metal accretion and mergers}
\label{sec:accretion}
The metal mass obtained by a galaxy through accretion ($M_m^{\rm acc}$) can be expressed as
\begin{equation}
M_m^{\rm acc}(z) = Z_{\rm IGM}(z) M_g^{\rm acc}(z) = Z_{\rm IGM}(z) \frac{\Omega_b}{\Omega_m} M_h^{\rm acc}(z),
\end{equation}
where $Z_{\rm IGM}$ is the IGM metallicity and we have reasonably assumed that the accreted gas mass is related to the accreted dark matter mass through the cosmological baryon-to-dark matter ratio. Further, the metal mass gained through mergers ($M_{m}^{\rm mer}$) can be calculated as 
\begin{equation}
M_{m}^{\rm mer}(z) = \sum_{j=1}^N M_{m,j}(z+\Delta z),
\end{equation}
\label{eq:acc2}
where $M_{m,j}$ is the final metal mass brought in by each of the $N$ progenitors from the previous redshift step ($z+\Delta z$). This term is naturally zero for starting leaves that have no progenitors.

With this setup, at the beginning of a given snapshot, the initial metal mass and gas-phase metallicity are:
\begin{equation}
\begin{split}
M_m^{\rm i}(z) &= M_m^{\rm acc}(z) + M_m^{\rm mer}(z),\\
Z^{\rm i}(z)   &= \frac{M_m^{\rm i}(z)}{M_g^{\rm i}(z)}.
\end{split}
\end{equation}

\subsection{Galactic chemical evolution}
\label{sec:chem}
The total mass rate of enriched and un-enriched material that dying stars return into the ISM at time $t$ is \citep[e.g.][]{Matteucci2016}:
\begin{equation}
G(t) = \int_{m(t)}^{50\ \Msun} [m - M_R(m,Z)] {\rm SFR}(t - \tau_m) \phi(m)\,dm,
\label{eq:G}
\end{equation}
where $\tau_m$ is the stellar lifetime of a star of mass $m$  \citep{padovani1993}, $M_R$ is the remnant mass depending on the mass $m$ and metallicity $Z$ (Sec. \ref{sec_yields}), and $\phi(m)$ represents the IMF\footnote{The IMF is normalized such that $\int_{m_l}^{m_u} m \phi(m) dm = 1$, with lower and upper mass limits of $m_l=0.1\Msun$ and  $m_u=100\Msun$, respectively.}. Here, $SFR(t-\tau_m) \phi(m)$ is the birthrate of stars with mass $m$ at time $t-\tau_m$; multiplying this birthrate with the mass ejected by one star, $m-M_R(m,Z)$, yields the total mass rate ejected by stars of mass $m$ at time $t$.
Analogously, the time-dependent rate of the ejected metals can be expressed as \citep[e.g.,][]{Yates2013}:
\begin{equation}
e_Z(t) = \int_{m(t)}^{50\ \Msun} m_y(m,Z) {\rm SFR}(t - \tau_m) \phi(m)\,dm,
\label{eq:eZ}
\end{equation}
where $m_y(m,Z) = y(m,Z) + (m - M_R(m,Z))Z$ is the mass of metals returned to the ISM by a star of mass $m$ and metallicity $Z$. 

This term is composed of the mass and metallicity dependent yield, $y(m,Z)$, which is the metal mass newly formed and ejected into the ISM by stars with initial mass $m$ and metallicity $Z$ (Sec. \ref{sec_yields}). The second term, $(m - M_R)Z$, accounts for the mass of metals present at the formation of the star restored back into the ISM without any nuclear processing \citep{Tinsley1980}. 

We can then split the term $e_Z(t)$ above into two parts:
\begin{equation}
\begin{split}
e_Z(t) =  &\int_{0.85\ \Msun}^{50\ \Msun} m_y(m,Z) {\rm SFR}(t - \tau_m) \phi(m)\,dm,\\
+ \Gamma  &\int_{\tau(0.85\ \Msun)}^{\tau(8\ \Msun)} m^{\rm SNIa}_y {\rm SFR}(t - \tau) {\rm DTD}(\tau)\,d\tau.\\
\end{split}
\label{eq:gce_split}
\end{equation}
The first term accounts for the contribution from stellar winds and ejection from single massive stars exploding as SNII. The second term represents the contribution from SNIa, where the stellar yield, $m^{\rm SNIa}_y$, is assumed to be independent of mass and metallicity as detailed in Sec. \ref{sec_yields} \citep[see also][]{Cappellaro2007,kobayashi_leung_nomoto2020}. Further, $\Gamma$ is defined as $\Gamma = A' k$, where $k = \int_{m_l}^{m_u} \phi(m) dm$ is the number of stars in a 1 $\Msun$ simple stellar population. 
Finally the $A'$ term accounts for the fraction of stars from the whole IMF that are SNIa progenitors such that \citep{Arrigoni2010,Yates2013}:
\begin{equation}
A' = A \cdot f_{3-16}.
\end{equation}
Here, $A$ represents the fraction of objects in the mass range 3-16 $\Msun$ that are SNIa progenitors and $f_{3-16}$ is the fraction of all objects in the IMF that have masses between 3 and 16 $\Msun$. For the IMF used in this work, $k = 2.8461$ and $f_{3-16} = 0.0258$. The assumed value for $A$ in this work is $A = 0.03$, i.e. 3 per cent of stars in the 3-16 $\Msun$ mass range are SNIa progenitors \citep{Yates2013}. The function DTD($\tau$) in the second term of Eqn. \ref{eq:gce_split} is an analytic power-law SNIa delay-time distribution (DTD) parametrized as \citep{Maoz2012,Yates2013}:
\begin{equation}
\text{DTD}(\tau) = a \left(\frac{\tau}{\text{Gyr}}\right)^{-1.12},
\end{equation}
where $a$ is assumed to be 0.15242 Gyr$^{-1}$. This DTD formalism allows us to not have to make additional assumptions about the progenitor type of SNIa, the binary mass function, secondary mass fraction distribution or binary lifetimes \citep{Yates2013}. Note that in this formalism the minimum delay-time of SNIa is determined from the lifetime of $8 M_\odot$ stars, which is about 30 Myrs. The adopted DTD is very similar to that in \citet{kobayashi_nomoto2009} around solar metallicity.

For the gas mass returned to the ISM and described by $G(t)$ in Eqn. \ref{eq:G}, we can use the same formalism in order to split the various contributions such that:
\begin{equation}
\begin{split}
G(t) =    &\int_{0.85\ \Msun}^{50\ \Msun} [m - M_R(m,Z)] {\rm SFR}(t - \tau_m) \phi(m)\,dm,\\
+ \Gamma  &\int_{\tau(0.85\ \Msun)}^{\tau(8\ \Msun)} m^{\rm SNIa} {\rm SFR}(t - \tau) \text{DTD}(\tau)\,d\tau,\\
\end{split}
\label{eq:return_split}
\end{equation}
where we take into account the various mass- and metallicity-dependent remnant masses $M_R(m,Z)$. Further, $m^{\rm SNIa}$ represents the amount of mass returned to the ISM by SNIa \citep[Sec. \ref{sec_yields},][]{snaith2015}. 

Note that in Eqns. \ref{eq:eZ}, \ref{eq:gce_split} and \ref{eq:return_split}, $Z$ depends both on time and the stellar lifetime i.e. $Z = Z(t - \tau_m)$, in order to account for the metallicity of stars at the time of their formation. 
In this work, we set the upper mass for chemical enrichment to 50 $\Msun$ because more massive stars do not explode as core-collapse supernovae and instead collapse to black holes without producing any heavy elements except for C and N \citep{kobayashi2020}.

\subsection{Stellar yields}
\label{sec_yields}
The stellar metal yields, $y(m,Z)$, can be obtained from stellar evolution and nucleosynthesis calculations for different ranges of stellar masses and initial metallicities. In this work, we adopt the latest yield tables from \citet{kobayashi2020}, which include all mass ranges of stars. These include low/intermediate-mass ($\sim 1-8 \Msun$) stars producing and losing metals through asymptotic giant branch (AGB)-phase stellar winds and core-collapse supernova explosions ($10-50 \Msun$) that produce the majority of metals. Among core-collapse supernovae, we assume half of $20-50 \Msun$ stars to form hypernovae (observed as Type Ibc supernovae) and the rest as SNII, which gives the best match to the observed elemental abundances in the solar neighbourhood. Among SNII, stars of $30-50 \Msun$ do not produce iron-peak elements, called failed supernovae. As reference, the return fraction and net yields are $R(Z=0)=0.36, R(Z=0.02)=0.41, y(Z=0)=0.018$, and $y(Z=0.02)=0.017$, depending on the metallicity (see Table 3 of \citealt{kobayashi2020} for the values with Kroupa IMF). The adopted solar metallicity is $Z_\odot=0.0144$ \citep{asplund2009}, and the solar oxygen abundance is $A_\odot({\rm O})=8.76$ \citep[for details see][]{kobayashi2020}.

For SNIa, we use a value of $m^{\rm SNIa}=1.37 \Msun$ independent of metallicity, and adopt $m^{\rm SNIa}_y$ from the spherical symmetric calculation named \quotes{W7} of \citet{Thielemann2003} for non-radioactive species.
At the solar metallicity, these yields are similar to those from more recent 2D calculations by \citet{kobayashi_leung_nomoto2020} for near-Chandrasekhar mass progenitors, except for Ni.

Oxygen yields are larger for more massive SN (see Fig. 4 of \citealt{nomoto2013}), while the mass dependence of iron yields is much smaller. Because of this difference there is a small decrease of [O/Fe] toward higher metallicities. In the galactic chemical evolution model for the solar neighbourhood, around [Fe/H] $\sim 1$, the [O/Fe] ratio sharply decreases due to the delayed enrichment from SNIa. This evolutionary trend is in an excellent agreement with the observed [O/Fe] ratios of nearby stars in the solar neighbourhood, if the \citet{kobayashi_nomoto2009} DTD model is used. While the DTD adopted in this paper gives an inferior fit to the observational data (see Fig.15 of \citealt{kobayashi_leung_nomoto2020}), this does not cause any significant difference for the high-redshift galaxies studied in this paper.

\subsection{Gas and metal mass ejection}
At each redshift-step, the gas and metal masses after accretion, merging, star formation and gas restoration/metal production are:
\begin{equation}
\begin{split}
M'_g(z) &= M_g^{\rm i}(z) - M_*^{\rm new}(z) + G(z) \Delta t,\\
M'_m(z) &= M_m^{\rm i}(z) - Z^{\rm i}(z) M_*^{\rm new}(z) + e_Z(z) \Delta t.
\end{split}
\label{eq:aftersf}
\end{equation}

Consequently, the metallicity at this stage can be expressed as:
\begin{equation}
Z'(z) = \frac{M'_m(z)}{M'_g(z)}.
\end{equation}

Assuming perfect mixing between gas and metals, the gas and metal mass ejected from the ISM (into the IGM) can be written as:
\begin{eqnarray}
\label{eq:ejection}
M_g^{\rm ej}(z) &= &[M_g^{\rm i}(z) - M_*^{\rm new}(z) + G(z) \Delta t] \frac{\feff}{\fej},\\
M_m^{\rm ej}(z) &= & Z'(z) M_g^{\rm ej}(z)
\label{eq:ejection_met}
\end{eqnarray}

As seen from the the terms $G(z) \Delta t$ and $e_Z(z) \Delta t$, we assume that newly formed metals and the returned gas are instantaneously mixed into the ISM before being ejected.

The {\it average} IGM metallicity, $Z_{\rm IGM}$, is assumed to be zero at the beginning of our simulation. For subsequent snapshots, the ejected metal mass reported in Eqn. \ref{eq:ejection_met} is used to compute its value as
\begin{equation}
Z_{\rm IGM}(z)  = \frac{1}{M_{g,{\rm tot(z)}}}\sum_{z'}^z \sum_{i=1}^N M_{m,i}^{\rm ej}(z'),
\label{eqn_zigm}
\end{equation}
where the RHS is the sum of the metal mass ejected over all galaxies ($N$) and simulation snapshots ($z' \geq z$) divided by the total gas mass in the simulation box at redshift $z$. Given the complexity of calculating $Z_{\rm IGM}$ on the full grid used for reionization, we defer this calculation to a future work.

\subsection{Evaluating gas and metal mass for each simulation snapshot}
\label{sec:gasmet}
The basic equation for the chemical evolution of the gas mass can be written as \citep[e.g.,][]{Matteucci2016}:
\begin{equation}
\frac{dM_g}{dt} = -{\rm SFR}(t) + G(t) - \frac{dM_g^{\rm ej}}{dt},
\end{equation}
that in our formalism translates into:
\begin{equation}
M_g^{\rm f}(z) = M_g^{\rm i}(z) - M_*^{\rm new}(z) + G(z) \Delta t - M_g^{\rm ej}(z),
\label{eq:return_gas}
\end{equation}
where the superscript \quotes{f} stands for \quotes{final} and indicates that the quantity is evaluated at the end of the redshift-step. Hence, Eqn. \ref{eq:return_gas} leads to:
\begin{equation}
M_g^{\rm f}(z) = [M_g^{\rm i}(z) - M_*^{\rm new}(z) + G(z) \Delta t] \left(1-\frac{\feff}{\fej}\right)
\end{equation}

The equivalent equation for the metal mass evolution is:
\begin{equation}
\frac{dM_m}{dt} = -Z \ {\rm SFR}(t) + e_Z(t) - \frac{dM_m^{\rm ej}}{dt},
\end{equation}
that again, in our formalism, leads to:
\begin{equation}
M_m^{\rm f}(z) = M_m^{\rm i}(z) - Z^{\rm i}(z) M_*^{\rm new} + e_Z(z) \Delta t - M_m^{\rm ej}(z),
\label{eq:final_met}
\end{equation}
and, consequently:
\begin{equation}
\begin{split}
M_m^{\rm f}(z) &= M_m^{\rm i}(z) - Z^{\rm i}(z) M_*^{\rm new}(z) + e_Z(z) \Delta t\\
               &-Z'(z)[M_g^{\rm i}(z) - M_*^{\rm new}(z) + G(z) \Delta t] \frac{\feff}{\fej}.\\
\end{split}
\end{equation}
This yields a final metallicity of
\begin{equation}
Z^{\rm f}(z) = \frac{M_m^{\rm f}(z)}{M_g^{\rm f}(z)}
\end{equation}

Given that in our simulation we are able to track the evolution of the Oxygen mass, we can express galaxy metallicities as 12 + log(O/H) using the following relation :
\begin{equation}
\text{12+log(O/H)} = 12 + \text{log}\left( \frac{M_{\rm O}^{\rm f}}{M_g^{\rm f}} \frac{m_{\rm H}}{m_{\rm O}X}\right)
\end{equation}
where $M_{\rm O}^f$ is the Oxygen mass, $m_{\rm H}$ and $m_{\rm O}$ are the atomic masses of Hydrogen and Oxygen, respectively, and $X$ represents the Hydrogen abundance. We have assumed $m_{\rm H} = 1.0079$, $m_{\rm O} = 15.999$ and $X = 0.75$. 

\begin{figure*}
	\centering
	\includegraphics[width=1.0\linewidth]{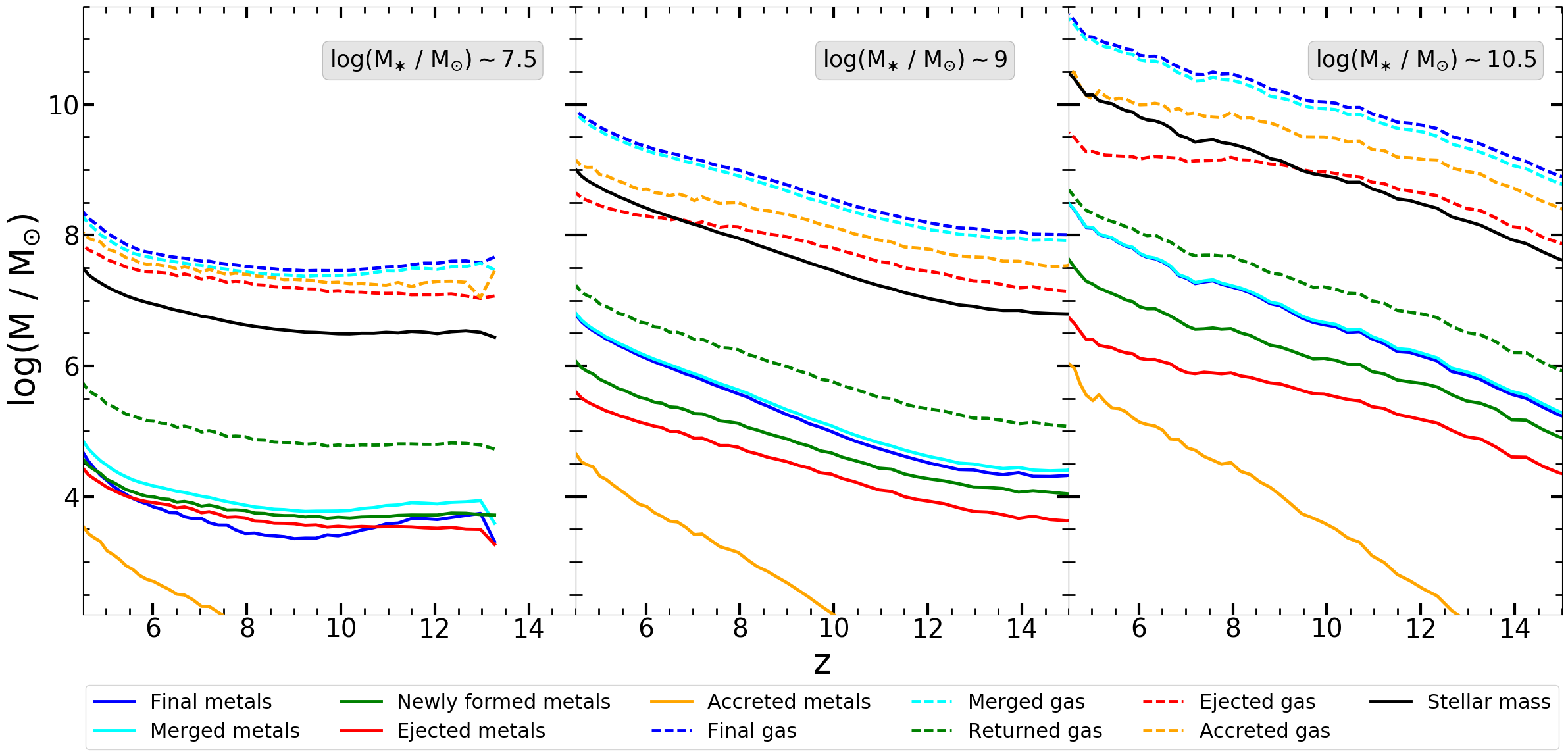}
	\caption{The average mass assembly history as a function of redshift for galaxies of three different mass ranges at $z \sim 4.5$, as marked in each panel. In each panel, solid and dashed lines show the results for the metal and gas masses, respectively. As also marked below the figure, the different lines show: the final mass (\textit{blue}), merged mass (\textit{cyan}), newly formed mass (\textit{green}), ejected mass (\textit{red}) and accreted mass (\textit{orange}); the solid black line shows the average stellar mass assembled. For clarity of visualisation, we only present results for the \textit{Photoionization} model, given that the \textit{Jeans mass} model leads to very similar values (see text in Sec. \ref{sec_assembly} for details).}
	\label{fig:assembly_p}
\end{figure*}

\section{The gas and metal content of early galaxies}   
\label{sec3}
We now discuss the key processes that drive the assembly of the gas and metal contents of early galaxies in Sec. \ref{sec_assembly} and describe the mass loading of gas and metals in Sec. \ref{sec:loading}. We primarily focus on the results for the \textit{Photoionization} model (with minimal UV feedback) and highlight the impact of (maximal) UV feedback through the \textit{Jeans mass} model where appropriate. Finally, for simplicity, the (dark matter, gas, stellar and metal) masses gained from {\it all} progenitors at the previous redshift-step are termed as \textit{merged} mass in what follows.

\subsection{Assembling the gas and metal content of early galaxies}
\label{sec_assembly}
In this section, we study the impact of the different physical processes discussed above in assembling the gas, stellar and metal contents of high-$z$ galaxies. As noted above, the gas and metal masses accreted or returned/produced in one redshift-step are counted as the merged gas and metal mass in the next step. 

We start by looking at the average mass assembly of low-mass galaxies ($M_* \sim 10^{7.5} \msun$ at $z \sim 4.5$) as shown in the \textit{left} panel of Fig. \ref{fig:assembly_p}. On average, these galaxies are hosted by halos of mass $M_h \sim 10^{9.8}~ (10^{10})\Msun$ at $z \sim 10 ~ (4.5)$ and start assembling their mass at $z \sim 13.5$. The flatness of their assembly stems from the fact that the majority of the newly-forming progenitors of such systems have low-masses at any redshift and therefore form stars in the \quotes{feedback-limited} regime where $\feff = \fej$. It is only at $z \lsim 6$ that their progenitors start growing significantly in terms of their stellar and gas content. 

The gas mass assembly of such galaxies is dominated by mergers bringing in $\sim 90\%$ of the final gas mass at all redshifts; most of this gas mass is contributed by the major progenitor. Smooth-accretion of IGM gas is the second most dominant process, contributing $\sim 60\%$ of the merged mass. Given the low-halo masses associated with the progenitors of such objects, ejection is about $40\%$ as important as mergers in determining the gas mass. The returned fraction ($\sim 0.02$ of the stellar mass) plays a negligible role, being about 2.5 orders of magnitude lower than the merged gas mass. In terms of metals, mergers are again responsible for assembling the bulk ($\sim 90\%$) of the total metal content at any redshift. This is followed by the metal enrichment from star formation contributing $\sim 70\%$ and ejection removing $\sim 40\%$ of the merged metal mass. Finally, given the relatively low value of the IGM metallicity ($Z_{\rm IGM}\sim 10^{-3} Z_{\odot}$ only by $z \sim 4.5$; see Sec. \ref{sec:igm_met}), smooth-accretion from the IGM contributes only about $5\%$ to the total metal mass, even by $z \sim 4.5$. 

We then show the assembly of intermediate mass galaxies ($M_* \sim 10^9 \msun$ at $z \sim 4.5$) in the {\it middle} panel of Fig. \ref{fig:assembly_p}. At $z = 10$ ($z = 4.5$) these galaxies are hosted in halos with masses $M_h \sim 10^{11}~(10^{11.2}) \msun$. As a result of their larger host halo masses, the progenitors of such galaxies are effectively unaffected by SNII feedback, allowing them to retain most, if not all, of their gas mass. This results in the gas (and stellar) mass growing monotonically with decreasing redshift between $z \sim 15-4.5$.  Mergers still dominate the gas mass assembly, contributing $\sim 90\%$ to the total gas mass at any redshift. The contribution from accretion decreases slightly with decreasing redshift from about 30\% of the merged gas mass at $z \sim 15$ to about 20\% by $z \sim 4.5$. The importance of ejection also decreases with decreasing redshift as the halo potential increases: while the ejected mass is about 15\% of the merged mass at $z \sim 15$, this decreases to $\sim 5\%$ by $z \sim 4.5$. Finally, the returned gas fraction is still $\sim 2.5$ orders of magnitude less than the merged mass. The metal-mass assembly is again dominated by mergers which bring in about 90\% of the metal mass at all redshifts, as might be expected from the gas-mass assembly. Self enrichment from star formation is about 50\% as important as mergers, with ejection removing $\sim 20\%$ of the merged metal mass at all redshifts; accreted metals again contribute a negligible $0.7\%$ to the total metal mass even by $z \sim 4.5$.

\begin{figure*}
	\centering
	\includegraphics[width=1.0\linewidth]{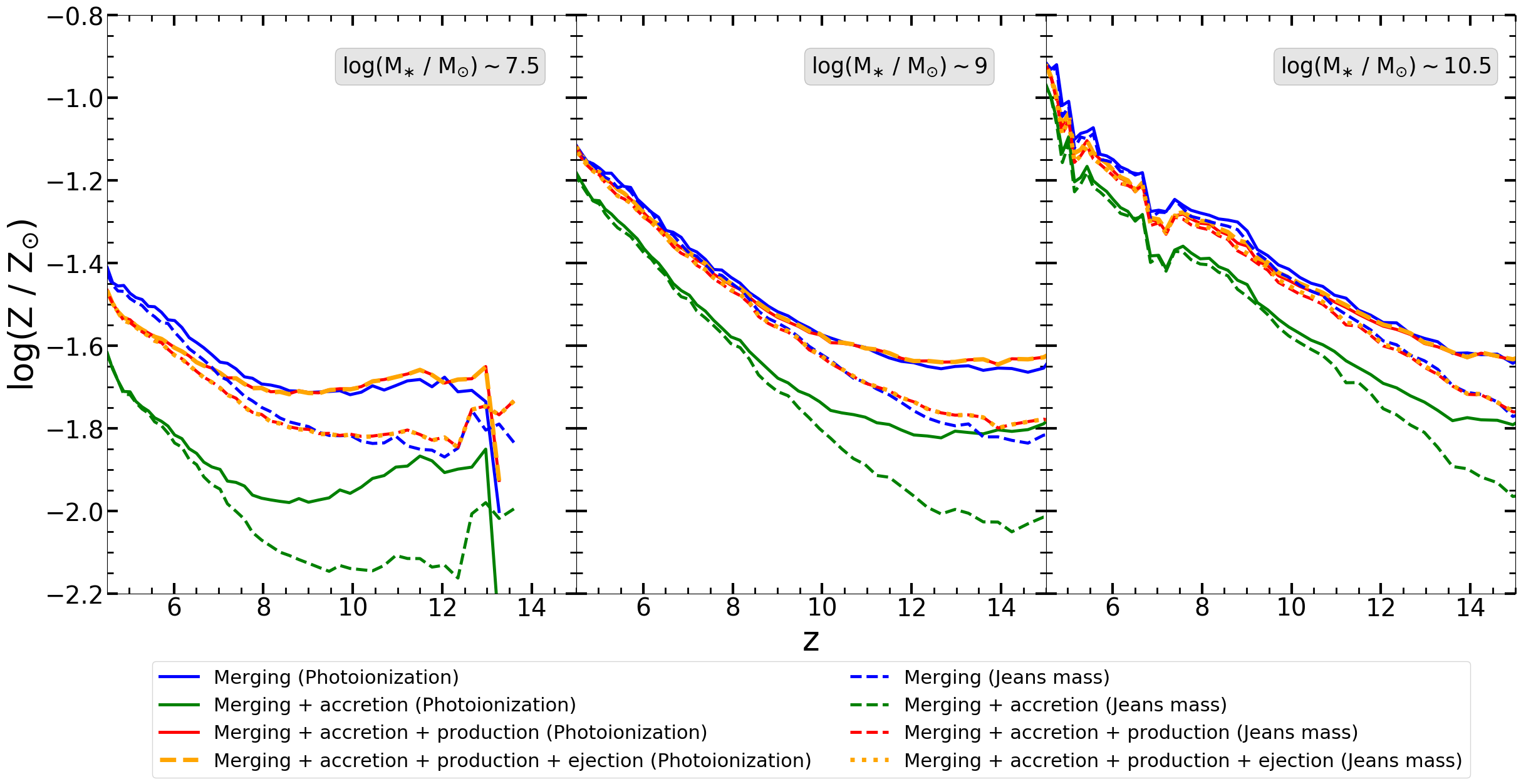}
	\caption{The average metallicity assembly history as a function of redshift for galaxies of three different mass ranges at $z \sim 4.5$, as marked in each panel. In each panel, solid and dashed lines show the results for the {\it Photionization} and {\it Jeans mass} models, respectively. As also marked below the figure, the different lines show the metallicity when including one process at a time at the previous redshift-step: merging only (\textit{blue}), merging + accretion (\textit{green}), merging + accretion + metal production (\textit{red}), merging + accretion + metal production + ejection (\textit{orange}).}
	\label{fig:assembly_Z}
\end{figure*}

The situation for the most massive galaxies ($M_* \sim 10^{10.5} \msun$ at $z \sim 4.5$), reported in the \textit{right} panel of Fig. \ref{fig:assembly_p}, is qualitatively the same as the intermediate mass bin. Hosted in halos of $M_h \sim 10^{12.4}\Msun$ at $z \sim 4.5$, the progenitors of such halos are mostly unaffected by SNII feedback. Indeed, these galaxies show the fastest buildup of both the gas and stellar mass, especially at earlier times. While mergers still dominate in assembling the bulk ($\sim 90\%$) of both the gas and metal mass, the key differences with respect to intermediate mass-halos is that, as a result of their faster growing halo masses, the impact of both accretion and ejection decreases faster for both the gas and metal masses with decreasing redshift. Quantitatively, accretion and ejection are $\sim 40\%$ and $\sim 10\%$ as important as mergers at $z \sim 15$ which decreases to $\sim 15\%$ and $\sim 2\%$ by $z\sim 4.5$, for both the gas and metal mass\footnote{Based on analytic merger tree arguments, the host halos of the low, intermediate and high mass galaxies studied here would evolve to to have masses $\sim 10^{10.25}, 10^{12}$ and $10^{14.6}\msun$ by $z \sim 0$}. 

We now briefly note the differences in the gas and metal assembly between the \textit{Photoionization} and \textit{Jeans mass} models. As might be expected, the stronger instantaneous radiative feedback in the {\it Jeans mass} model leads to a larger reduction in the gas mass (and hence the star formation in low-mass galaxies) at early times as compared to the {\it Photoionization} model \citep{astraeus1}. While the assemblies converge at lower-redshifts, the major differences arise at $z \gsim 10$ where the (accreted and merged) gas masses are lower by about 0.1 dex in the {\it Jeans mass} model; the corresponding difference in the metal mass is of the order of 0.3 dex given the strong dependence of metal production on the star formation rate.

\begin{figure*}
	\centering
	\includegraphics[width=0.482\linewidth]{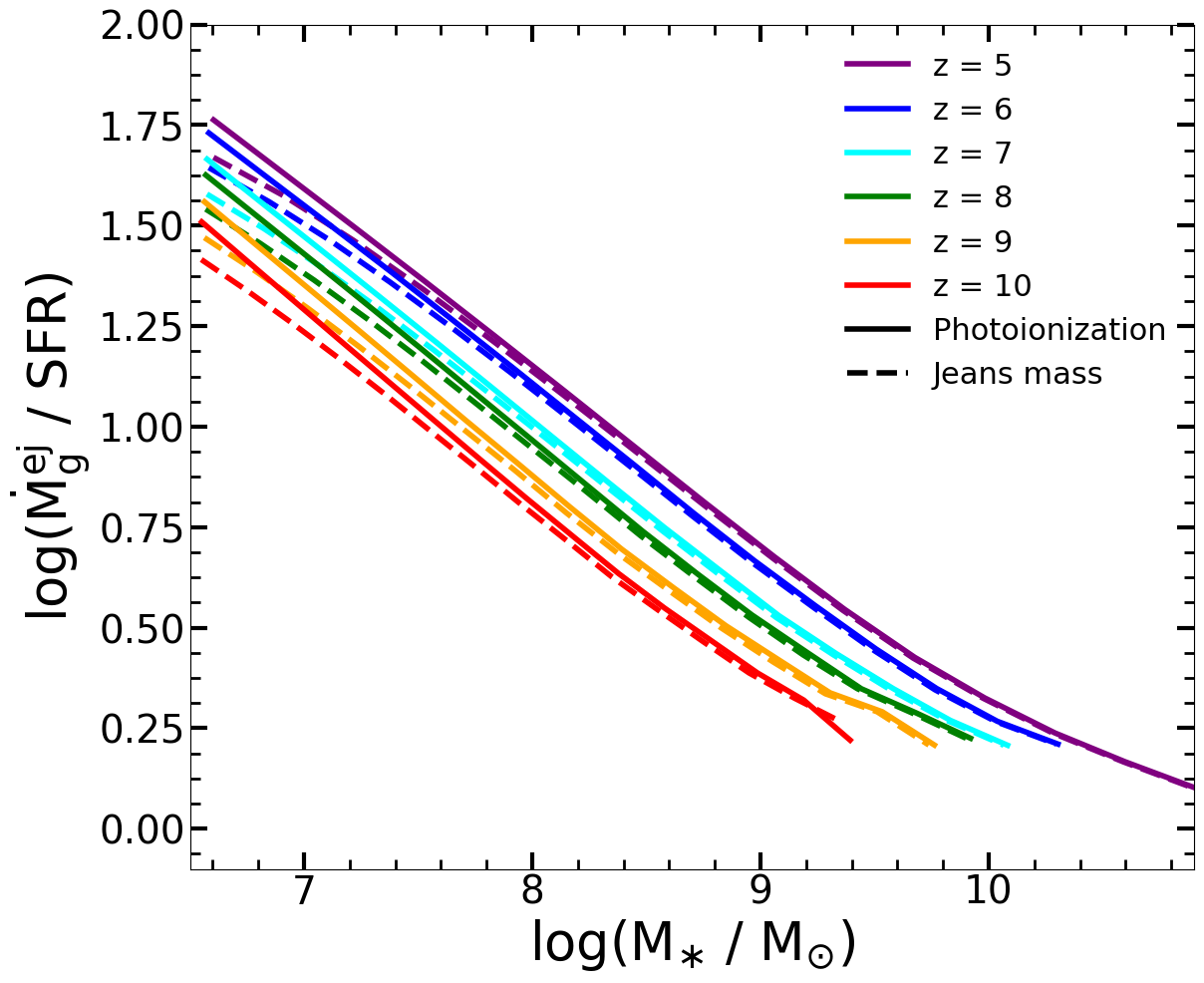}
	\includegraphics[width=0.482\linewidth]{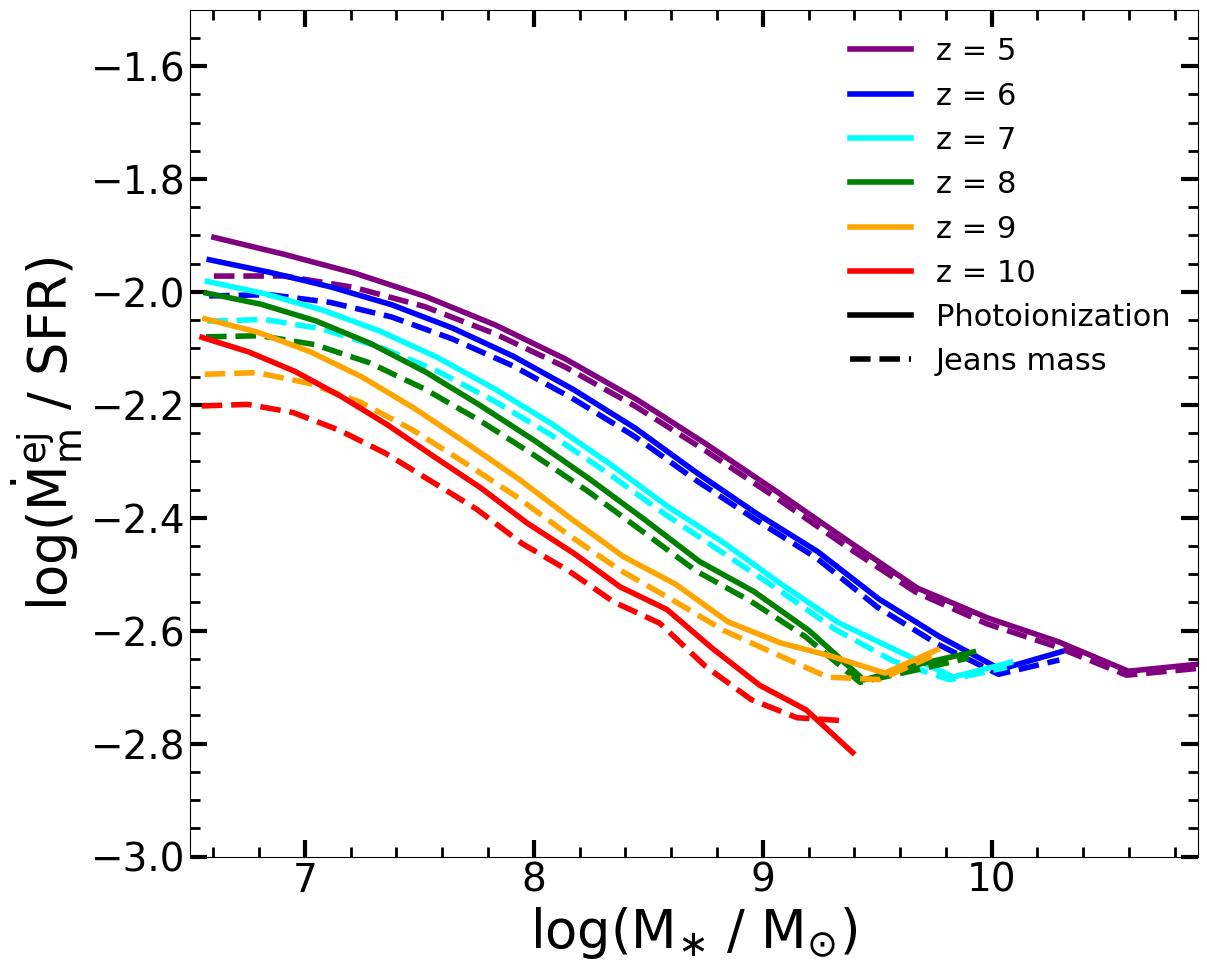}
	\caption{As a function of the stellar mass, for redshifts $z \sim 5-10$ as marked, we show: the average gas mass-loading factor, ${\rm log} ~ \eta_g = {\rm log}(\dot M_g^{ej}/{\rm SFR})$, in the \textit{left panel} and the average metal mass-loading factor, ${\rm log} ~ \eta_m = {\rm log}(\dot M_m^{ej}/{\rm SFR})$, in the \textit{right panel}. Solid and dashed lines show results for the \textit{Photoionization} and \textit{Jeans mass} models, respectively. }
	\label{fig:eta}
\end{figure*}

In Fig. \ref{fig:assembly_Z}, we show the average history of the gas-phase metallicity both for the \textit{Photoionization} and \textit{Jeans mass} models for the different physical processes considered in \textlcsc{ASTRAEUS}: merging, accretion, metal enrichment and gas return due to star formation and ejection of gas and metals due to SN feedback. At each redshift step shown, each of the processes is added one by one in order to show its impact. We start our discussion with the {\it Photoionization} model. Here, for low-mass halos (with $M_* \sim 10^{7.5} \msun$ at $z \sim 4.5$), we find that only considering the metals and gas brought in by mergers from the previous redshift-step results in the metallicity increasing by a factor of 2 from about $2\%~\zsun$ at $z \sim 13$ to $\sim 4\%$ by $z \sim 4.5$. The metallicity build-up shows a steep increase below $z \sim 8$, mirroring the increase in both the gas and metal masses. This is driven by the progenitors of such low-mass systems building up their potentials, allowing them to hold on to a larger fraction of their baryonic mass. Adding accretion has the immediate effect of a {\it dilution} of metals inside galaxies, since the smoothly-accreted gas mass is at least 4.5 orders of magnitude larger than the accreted metal mass as shown in Fig. \ref{fig:assembly_p}. Indeed, as seen from Fig. \ref{fig:assembly_p}, the accreted gas mass increases at $z \lsim 8$ which naturally leads to a larger drop in the metallicity due to dilution by about 0.2 dex. Given the metal-produced to gas-returned ratio of 10\% (see Fig. \ref{fig:assembly_p}), including self-enrichment of metals from star formation compensates for the effect of dilution at $z \gsim 8$ where the metallicity goes back to (or even slightly above) the values considering mergers only. However, at lower redshifts, enrichment is unable to fully compensate for the influx of metal-poor accreted gas resulting in a metallicity value that is about 0.15 dex below that from mergers only. Finally, given that we assume perfect mixing of gas and metals, outflows have no impact on the gas-phase metallicity. 

We then study the effect of strong and instantaneous reionization feedback on the metallicity of these low-mass galaxies through the {\it Jeans mass} model. In this model, the gas (and hence star formation rates and the corresponding metal enrichment) of the progenitors of such low-mass halos are severely suppressed as compared to the {\it Photoionization} model, especially at high-redshifts ($z \gsim 8$); the effects of feedback become less prominent as these galaxies build-up their potentials steeply at lower-redshifts. While the metallicity from mergers is $\sim$ 0.2 dex lower in the {\it Jeans mass} model at $z \sim 8$, it converges towards the results from the {\it Photoionization model} at lower-$z$. When accretion is added, the metallicity decreases by $\sim 0.3$ dex. As expected, including self-enrichment increases the metallicity value to the merger-only scenario with ejection again leaving the metallicity unaffected. Including all these processes, the final metallicity is about $0.18$ dex lower in the {\it Jeans mass} model as compared to the {\it Photoionization model} at $z \gsim 8$, with the results converging by $z \sim 5.5$. 

Qualitatively, the results are very similar for the more massive galaxies shown in the middle and right-most panels of Fig. \ref{fig:assembly_Z}. Galaxies with $M_* \sim 10^9 \msun$ by $z \sim 4.5$ show a smoother metallicity build-up given the larger progenitors of such systems. In the {\it Photoionization} model, considering gas and metals from mergers only, the metallicity value increases from $\sim 2-8\%~\zsun$ between $z \sim 15$ and $4.5$. Including smooth-accretion leads to a drop of about 0.15 dex at $z \sim 15$. As the progenitors build their mass, the impact of accretion decreases such that dilution only decreases the metallicity by about 0.1 dex by $z \sim 4.5$. Self-enrichment increases the metallicity values, essentially bringing them back to the ``mergers-only" scenario. In terms of strong reionization feedback, as expected, the {\it Jeans mass} model shows the largest decrease in the final metallicity ($\sim 0.2$ dex) at $z \gsim 12$. As the progenitors of such systems exceed the halo mass ($\sim 10^{9.5} \msun$) impacted by even this strong feedback, these metallicity values come into accord with the {\it Photoionization model} by $z \sim 8$. 

Finally, galaxies with $M_* \sim 10^{10}\msun$ at $z \sim 4.5$ show the smoothest metallicity evolution (from $2-14\%~\zsun$ between $z \sim 15$ and $4.5$). While the impact of accretion and production are the same as in the intermediate mass case discussed above, the key difference seen in this model is that the results from the {\it Photoionization} and {\it Jeans mass} models start converging at redshifts as high as $z \sim 10$. 

To summarise, while merging dominates the mass assembly of both the gas and metal contents of all the stellar masses studied  here at $z \gsim 4.5$, the importance of accretion, ejection and self-enrichment decrease with an increase in the halo mass. For metallicity, merging again plays the predominant role. While gas accretion from the IGM dilutes the metallicity, this is {\it effectively restored} by self-enrichment; in the perfect mixing scenario (between gas and metals) considered here, ejection has no impact on the metallicity. 

Finally, since we assume metals to be homogeneously distributed throughout the simulation box to determine the IGM metallicity (discussed further in Sec. \ref{sec:igm_met}), the metal-mass accreted from the IGM must be treated as a lower limit. Indeed, metals ejected by galaxies at a given time can be re-accreted at later times resulting in the accretion of gas that can be significantly metal-rich compared to the average IGM metallicity  \citep[e.g.][]{dave2011,muratov2017,oppenheimer2018}. This could be highly relevant for the metallicity of clustered lowest-mass systems that build up a significant part of their gas mass through IGM accretion. However, the importance of IGM-accreted gas derceases with an increase in the stellar (or halo) mass. This implies that IGM accretion of metal enriched gas would have only a limited impact on the metallicity of the intermediate to high-mass systems discussed above.
\subsection{Mass-loading factors for gas and metals}
\label{sec:loading}
We then study the mass-loading factors for gas ($\eta_g$) and metals ($\eta_m$) that are defined as the outflow rates of gas and metals per unit SFR, respectively, such that
\begin{eqnarray}
 \eta_g  & = & \frac{\dot M_g^{\rm ej}} {\rm SFR} = \frac{M_g^{\rm ej}}{M_*^{\rm new}}, \\
 \eta_m & = & \frac{\dot M_m^{\rm ej}} {\rm SFR} = \frac{M_m^{\rm ej}}{M_*^{\rm new}}. 
\end{eqnarray}

In order to gain an intuitive understanding of the behaviour of these mass loading factors as a function of stellar mass and redshift, we assume instantaneous SN feedback and the minimal {\it Photoionization} feedback. This is a reasonable assumption at $z \lsim 6.5$ where the time difference between consecutive redshift steps is larger than the 28 Myrs  required for all SNII from a given stellar population to explode within a given redshift-step. This results in 

\begin{equation}
 \eta_g  = \frac{M_g^{\rm i}(z) - M_*^{\rm new}(z) + G(z) \Delta t}{M_*^{\rm new}(z)} \frac{\feff}{\fej}.
\end{equation}
As seen from Fig. \ref{fig:assembly_p}, $G(z) \Delta t \ll M_g^{\rm i}$. Disregarding this term and recalling that $M_*^{\rm new}(z) = \feff M_g^{\rm i}(z)$ yields
\begin{equation}
\eta_g \approx \frac{1-\feff}{\fej},
\label{etag}
\end{equation}
where $\fej \ll 1$ \citep[see Fig. 1;][]{dayal2014}. Low-mass halos with $M_h \lsim 10^{9.1} ~ (10^{9.5}) \msun$ at $z \sim 10 ~ (5)$ form stars in the feedback-limited regime such that $\feff = \fej$; larger halos form stars at a constant efficiency of $\feff \sim 1\%$ which is much lower than $\fej$. In this case, one naturally expects $\eta_g$ to decrease with an increase in the halo (or stellar) mass irrespective of redshift; this is exactly the trend seen in the left panel of Fig. \ref{fig:eta}. For example, at $z \sim 5$, $\eta_g$ decreases from about $48.6$ to $4.8$ as $M_*$ increases from $10^7$ to $10^9\msun$ for the {\it Photoionization model}.

Further, galaxies of a given stellar mass are hosted by halos of very similar mass (to within 0.2 dex) at $z \sim 5-10$ leading to the $\eta_g-M_*$ slope being similar at all these redshifts. However, halos of a given mass correspond to deeper potential wells with increasing redshift. This leads to an increase in $\fej$ that naturally leads to a corresponding decrease in $\eta_g$. Indeed, for a given stellar mass, $\eta_g$ decreases by about 0.35 dex between $z \sim 5$ and 10. For example, at $z \sim 10$, $\eta_g$ decreases from a value of about $17.8$ to $2.1$ as $M_*$ increases from $10^7$ to $10^9\msun$. For the {\it Photoionization model}, we find the following average relation between $z \sim 5-10$ for stellar masses between $10^{7-10.4}\msun$  ($10^{7-9.5}\msun$) at $z \sim 5$ (10):

\be
\eta_g \approx 1.38 \bigg(\frac{M_*}{10^{10}\msun}\bigg)^{-0.43}.
\ee
Because the mass loading (normalized to $M_*$) factor $\eta_g$ is independent of metallicity, but depends on SFR, we expect a mass dependence for $\eta_g$ through the relation of SFR with $M_*$.
Although the index we find is somewhat steeper than the $\eta_g \sim M_*^{-0.33}$ dependence expected in case of momentum-driven winds found in previous theoretical calculations \citep[e.g.][]{finlator2008, dave2012, dayal2013, muratov2015}, it is in agreement with the power law values of $-0.29$ to $-0.45$ expected from fitting analytic models to 
observational data at $z \sim 0-3.5$ \citep[see e.g.][]{hunt2016b}. The steepness of our relation is almost to be expected given our implementation of SNII energy coupling to gas - while low-mass halos can lose all of their gas content, halos above $10^{9.5}\Msun$ are hardly affected by such feedback.

In terms of reionization feedback, both the slope and amplitude of the $\eta_g-M_*$ relation for the {\it Jeans mass} model converge to the {\it Photoionization} model for $M_* \gsim 10^8\msun$ galaxies (corresponding to $M_h \gsim 10^{10.2}\msun$) for all redshifts. This is because their progenitors were large enough to not be affected by reionization feedback. However for lower stellar masses, the {\it Jeans mass} model shows a shallower slope. Being severely suppressed in terms of their gas mass in this model, galaxies of a given (low) stellar mass are hosted by more massive halos (by about 0.4 dex) as compared to the {\it Photoionization} model. These larger potentials naturally result in slightly higher $\fej$ values, resulting in $\eta_g$ values that are lower by about 0.25 dex in the {\it Jeans mass} model. Averaged over $z \sim 5-10$, we find the following (slightly shallower) relation for the {\it Jeans mass} model for the entire galaxy population:
\be
\eta_g \approx 1.39 \bigg(\frac{M_*}{10^{10}\msun}\bigg)^{-0.41}.
\ee

We then discuss the mass loading for metals that, using Eqns. \ref{eq:ejection_met} and \ref{etag}, can be expressed as
\begin{equation}
\eta_m \approx Z' \frac{(1-\feff)}{\fej}.
\end{equation}
We remind the reader that $Z'$ is the ISM metallicity after mergers, dilution from inflows and enrichment from star formation but before ejection. As shown in the right panel of Fig. \ref{fig:eta}, the trends for $\eta_m$ are very analogous to the behaviour shown by $\eta_g$. For the {\it Photoionization} model, $\eta_m \propto M_*^{-0.27}$ for $M_* \gsim 10^7 \msun$ at all redshifts; $\eta_m$ decreases from a value of 0.01 to 0.004 as $M_*$ increases from $10^7$ to $10^9\msun$ at $z \sim 5$. Below this stellar mass, the slope of the $\eta_m - M_*$ relation flattens with decreasing redshift as the progenitors of such galaxies lose more and more of their metal mass with time. The amplitude of this relation decreases with increasing redshift due to the increasingly deep potentials of the host halos of galaxies of a given stellar mass. The stronger feedback in the {\it Jeans mass} model results in lower star formation rates and therefore a lower metal enrichment, that is most pronounced at $z \gsim 9$. By $z \sim 8$, the results of the {\it Jeans mass} and {\it Photoionization} models are in accord for $M_* \gsim 10^8 \msun$ galaxies. However, at $M_* \lsim 10^7 \msun$, galaxies in the {\it Jeans mass} model show an almost flat slope: this is because the larger gas suppression in the progenitors of these halos results in correspondingly lower star formation rates and hence a lower self enrichment. Most of the metals produced and ejected in these halos are therefore from star formation in the redshift-step under consideration resulting in a flattening of the slope.  

\begin{figure*}
	\centering
	\includegraphics[width=0.96\linewidth]{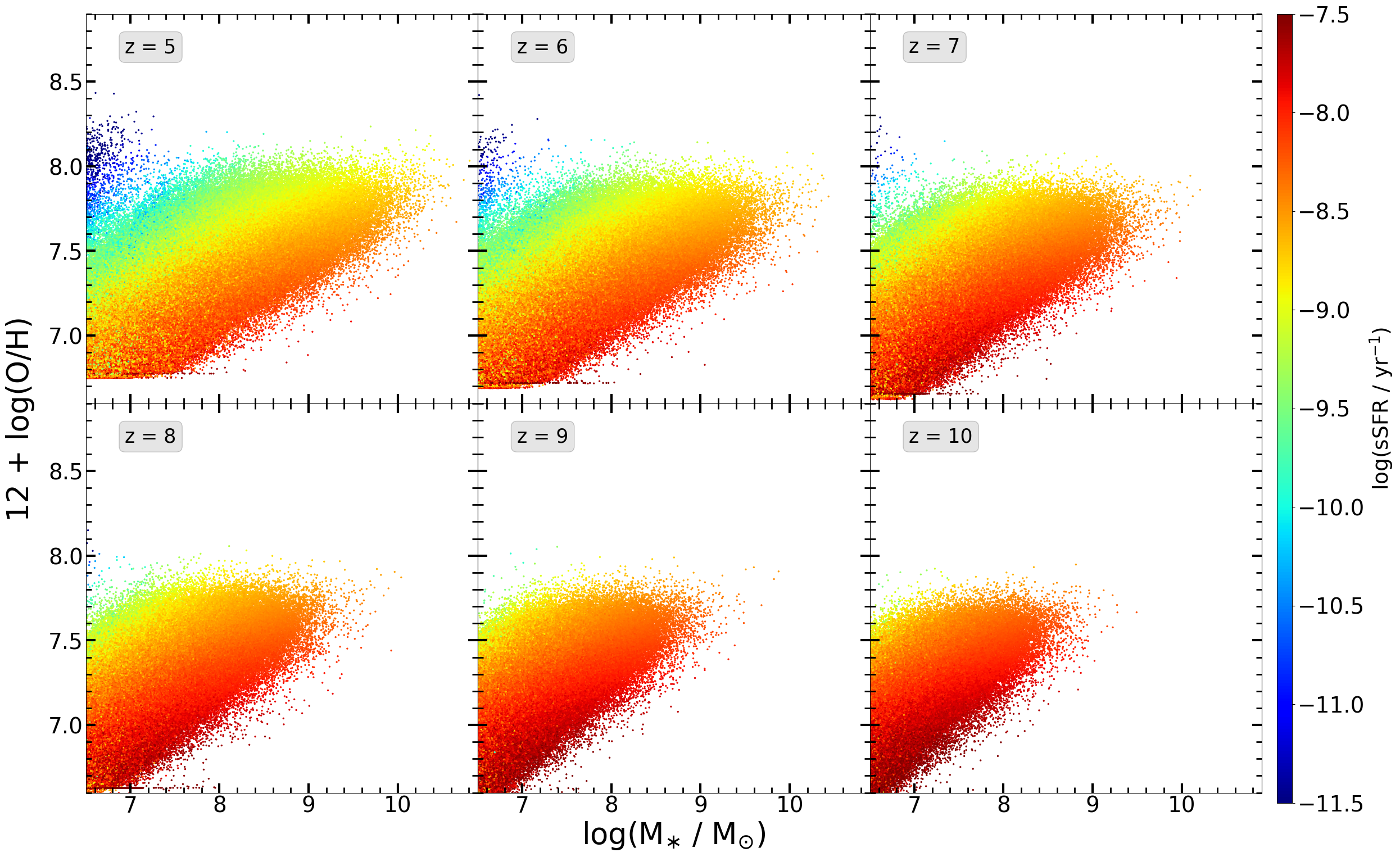}
	\caption{The mass-metallicity relation linking the (final) gas-phase metallicity, in units of $12+{\rm log~(O/H)}$, and the stellar mass for $z \sim 5-10$ as marked. In each panel, the results have been color-coded by the specific star formation rate. For clarity, we only show results for the \textit{Photoionization} model. }
	\label{fig:sSFR}
\end{figure*}
\section{The mass-metallicity relation and its redshift evolution}
We now discuss the dependence of the mass-metallicity relation on the specific SFR (sSFR) in Sec. \ref{sec:sfr} before showing the 3D fundamental metallicity relation at $z \sim 5-10$ in Sec. \ref{sec:HFPZ}. We then discuss the redshift evolution of the FMR in Sec. \ref{sec:zev} before ending with the metal enrichment of the IGM through cosmic time in Sec. \ref{sec:igm_met}.

\subsection{The metallicity dependence on the specific SFR}
\label{sec:sfr}
We start by showing the stellar mass-gas phase metallicity, color-coded by the specific star formation rate (sSFR) for $z \sim 5-10$ in Fig. \ref{fig:sSFR}; the same relation color-coded by the star formation rate (i.e. the FMR) is shown in Appendix \ref{app:fmr}. Some of the key trends emerging from this figure are: (i) at all redshifts the gas-phase metallicity shows a positive correlation with the stellar mass (as will be discussed in detail in Sec. \ref{sec:zev} that follows) and; (ii) independent of redshift, at a given stellar mass, the metallicity decreases with increasing sSFR. 

Starting at $z \sim 5$, our model tracks galaxies with $M_* \sim 10^{6.5}-10^{10.8}\msun$. We see that the lowest mass galaxies (with $M_* \lsim 10^{7.5}\msun$) show sSFRs that range over three orders of magnitude between $10^{-11.5}-10^{-8} \, {\rm yr^{-1}}$. This large range in sSFR is indicative of the variety of assembly histories through which such low-mass galaxies build up their gas mass. 
It is important to note that low-mass galaxies ($M_* \lsim 10^{7.5}\msun$) showing metallicity values of 12+log(O/H)$\gsim 7.6$ are outliers. Their metallicity values, that can be high as $40\%~\zsun$ (see the left-most panel in also Fig. \ref{fig:assembly_p}), require a combination of: (i) dry mergers that do not bring in any gas or metal mass; and (ii) low gas accretion rates which result in the gas mass being dominated by that returned to the ISM from star formation while the metal mass is primarily determined by self-enrichment from newly produced metals. The first effect is a natural result of our ``maximal supernova feedback" model where low-mass galaxies form stars at an efficiency that can unbind the rest of the gas. Observations of such low-mass galaxies with e.g. the JWST will be crucial in confirming such a scenario.

For this stellar mass bin, disregarding these outliers, the metallicity decreases from 12+log(O/H)$\sim 7.8$ to $6.8$ as the sSFR increases from $10^{-10.25}$ to $10^{-8} \, {\rm yr^{-1}}$. For a given stellar mass, a higher SFR implies a larger gas reservoir for star formation which is built-up both from mergers and accretion of metal-poor gas from the IGM. As seen from Fig. \ref{fig:assembly_Z}, this {\it dilution} of metallicity due to IGM accretion of metal poor gas and the associated higher amount of ejection play a key role in the metallicity decreasing with increasing sSFR for these low-mass halos. 

The range of sSFR narrows with increasing stellar mass. For example, galaxies with $M_* \sim 10^{9.5}\msun$ (with $M_h \sim 10^{11.2}\msun$) show sSFR between $10^{-9.5}-10^{-8.5} \, {\rm yr}^{-1}$. This is because, due to their larger potential wells, the progenitors of such systems are much less affected by SN feedback. As a result, the gas masses of these halos are quite similar, being of the order of 10\% of the halo mass \citep [see Fig. 8;][]{astraeus1}. Even for such high-mass galaxies, the metallicity shows a decrease (although of only 0.7 dex) from 12+log(O/H) $\sim$ 8 to 7.3 as the sSFR increases from $10^{-9.5}$ to $10^{-8.5}$ yr$^{-1}$. This smaller decrease can be explained by the fact that the progenitors of such galaxies were massive enough to retain most of the (star-formation produced and accreted) metals within their potential for a substantial part of their assembly history (see also middle panel of Fig. \ref{fig:assembly_Z}). 

These same trends persist at all redshifts. As expected, both the stellar mass and sSFR ranges shrink with increasing redshift. While the former is a direct result of hierarchical structure formation, the latter is driven by the fact that galaxies of a given stellar mass are hosted by slightly more massive halos with increasing redshifts. Given that these correspond to higher-$\sigma$ fluctuations, they have fewer generations of low-mass progenitors that are feedback-limited. 

Finally, at $z \sim 10$, for $M_* \sim 10^{7.5} \msun$ galaxies, the metallicity decreases from 7.75 to 6.75 as the sSFR increases from $10^{-9.5}-10^{-7.5} \, {\rm yr^{-1}}$. Despite the simulation volume, we only find a few galaxies with a stellar mass of $10^9\msun$ at this redshift. These show both very similar metallicities (12+log(O/H)$\sim 7.5-7.8$) and sSFR ($\sim 10^{-8.25}\, {\rm yr^{-1}}$) hinting at the similar assembly histories of such highly-biased objects.

\begin{figure*}
	\centering
	\includegraphics[width=1.0\linewidth]{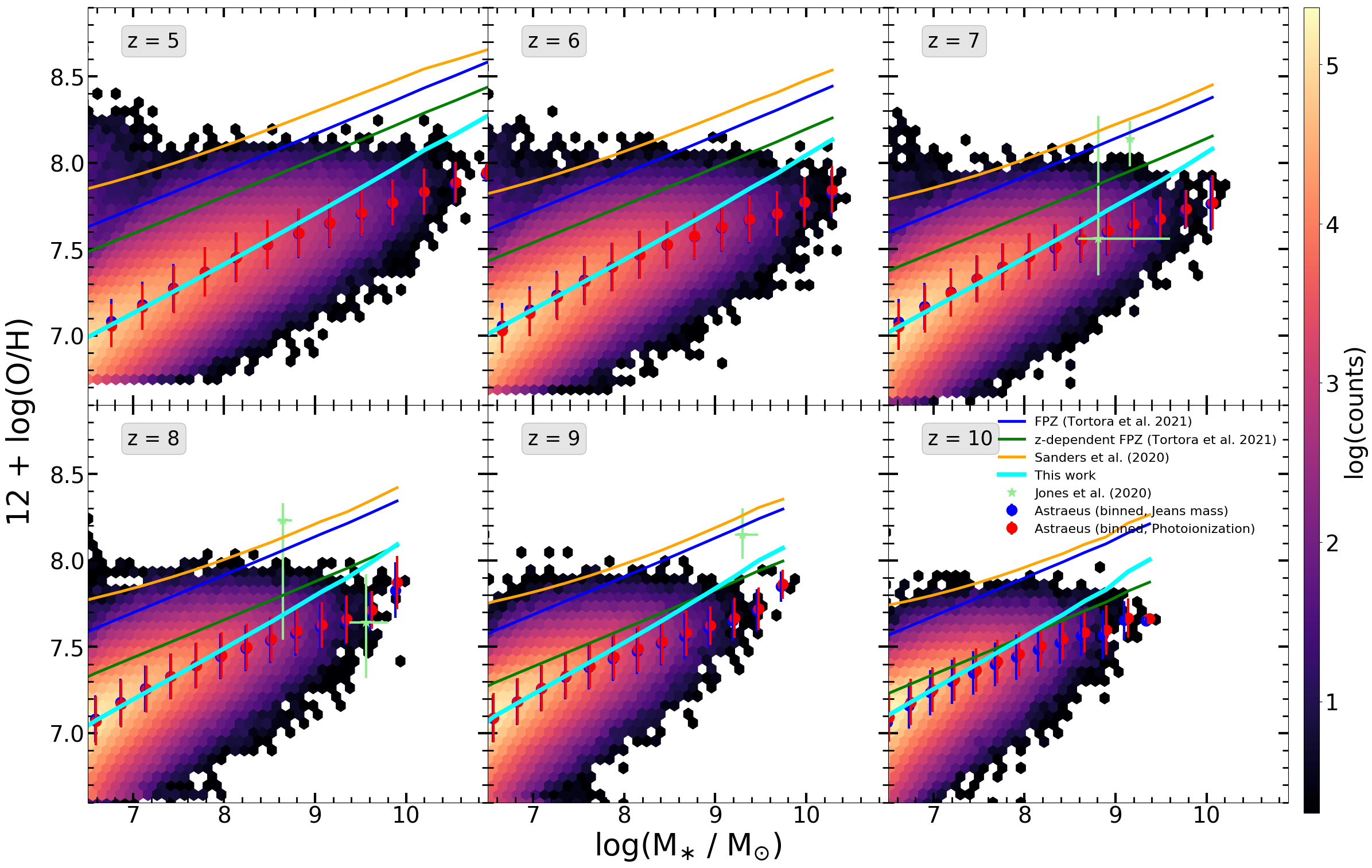}
	\caption{The panels show the mass-metallicity relation at $z \sim 5-10$, as marked in each panel. In each panel, the filled red and blue circles show the median metallicity in a given stellar mass bin for the \textit{Photoionization} and {\it Jeans mass} models, respectively; the error bars show the corresponding $1-\sigma$ dispersion. The cyan line shows the fundamental plane of metallicity (FPZ; see Sec. \ref{sec:HFPZ}) as obtained from this work for the {\it Photoionization} model. The solid blue (green) line show the observationally inferred parametrization of the redshift-independent (redshift-dependent) FPZ from 
	\citet{tortora2022}
	while the solid orange line shows the best-fit results from \citet{sanders2020}. Finally, green stars show results using the with direct-method constraints of \citet{jones2020}. All of these observational relations have been recalibrated to a $0.1-100\msun$ Salpeter IMF (Tortora and Sanders, private comm.).}
	\label{fig:mass_metallicity_hist_p}
\end{figure*}

\begin{figure}
	\centering
	\includegraphics[width=1.0\linewidth]{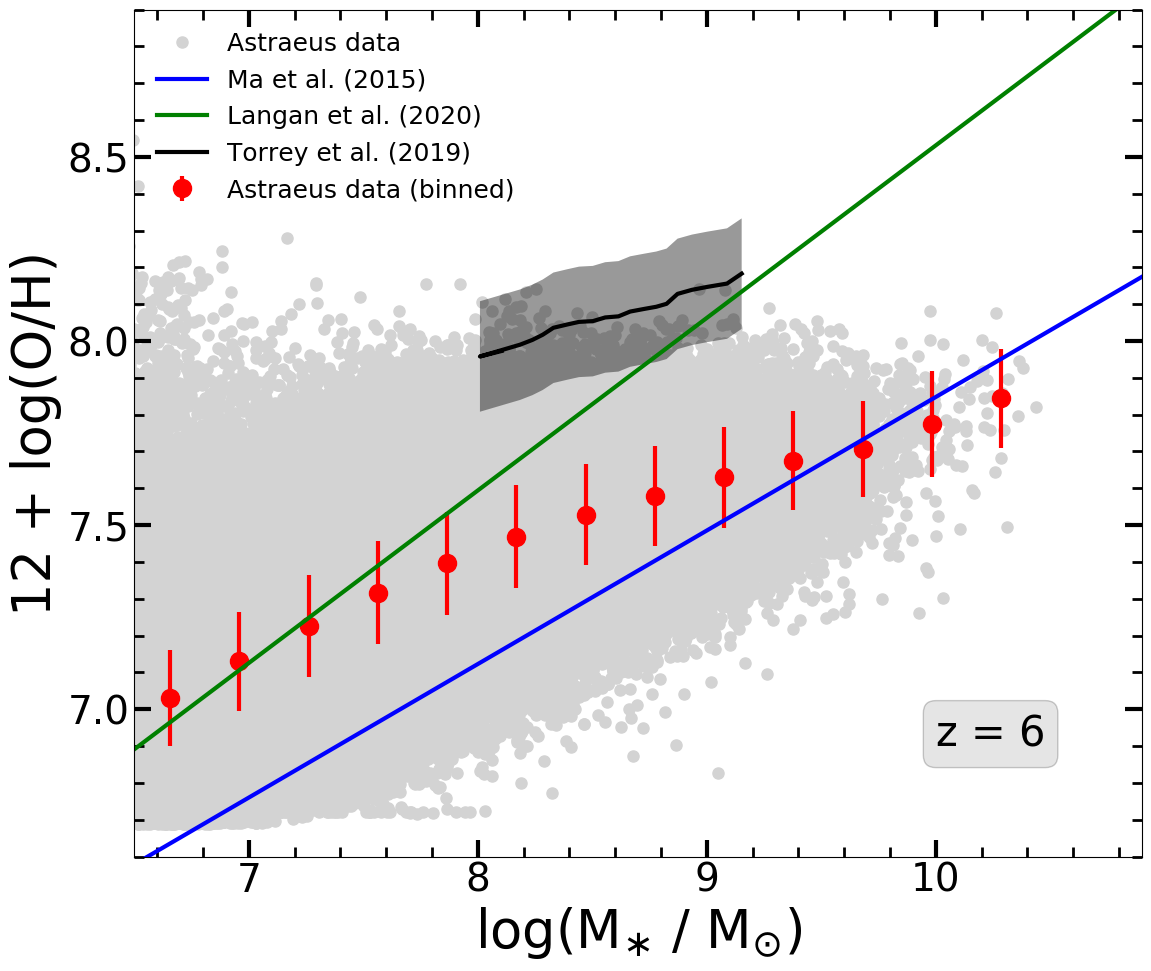}
	\caption{A comparison of the mass-metallicity relation at $z=6$ from different theoretical models. The gray points show results for all $z=6$ galaxies from \textlcsc{Astraeus} with the red points showing the median values. The blue line shows the fit predicted by the \textsc{fire} simulation \citep{ma2016} for galaxies with $M_* \sim 10^{3-9.1}\Msun$, the green line shows the linear fit from the \textsc{FirstLight} simulation \citet{langan2020} for galaxies with $M_* \sim 10^{6.25-9.25}\Msun$ and the black line shows the results from the \textsc{illustris tng} simulation \citep{torrey2019}, along with the $1-\sigma$ error bar. We have rescaled the metallicites from all these works to our solar value of 8.76.}
	\label{fig:comparison}
\end{figure}

\subsection{The emergence of the mass-metallicity relation and a fundamental plane of metallicity at high-redshifts}
\label{sec:HFPZ}
We now study the two-dimensional relation between the stellar mass and gas-phase metallicity in Fig. \ref{fig:mass_metallicity_hist_p}. The two key trends seen from this figure are that, firstly, the median metallicity scales with the stellar mass at all redshifts i.e. the {\it mass-metallicity relation (MZR) is already in place at $z \sim 10$ and persists at all the redshifts ($z \sim 5-10$) studied}.  This is driven by the fact that, as a result of their low SFRs low-mass galaxies produce fewer metals at any time which are more diluted due to smooth-accretion, a large fraction of which are progressively lost in outflows given their small potential wells. On the other hand, high-mass galaxies (and their progenitors) can produce more metals, a larger fraction of which are retained in the ISM and accreted into their successors, with dilution playing a decreasing role as the stellar mass increases. Secondly, at almost all redshifts, we see an upper envelope composed of a few heavily enriched outlying systems. Indeed, at $z \sim 5-8$, there are a handful of low-mass ($M_* \lsim 10^{7.5}\msun$) galaxies that show metallicity values as high as $40\%~\zsun$. As also noted in Sec. \ref{sec:sfr}, given their low-masses, such galaxies mostly assemble from dry mergers and have low gas accretion rates. This results in the gas mass being dominated by that returned to the ISM from exploding stars and the metal mass being primarily determined by self-enrichment from newly produced metals. Forthcoming observations of the metallicities of such systems with the JWST will be crucial in validating such a scenario. Thirdly, we do not see any evidence of metallicity saturation for high-mass galaxies. Low-redshift ($z \sim 0-3.5$) observational data-sets show that while the MZR is approximately linear up to $M_* \sim 10^{10.5} \msun$, it flattens for higher-mass galaxies \citep{tremonti2004,kewley2008,hunt2016,wyder2007} which could be driven by a balance between self-enrichment and dilution due to inflows of metal-poor IGM gas \citep{dayal2013}. However, within error bars, we do not see any clear indication of such a flattening at high stellar masses. Indeed, as seen from Sec. \ref{sec_assembly}, while accretion is certainly important in determining the gas content of such early galaxies, the decrease in metallicity due to such dilution is superseded by that from self-enrichment for the entire mass range modelled here (also see Fig. \ref{fig:assembly_Z}).

In terms of median metallicities, at $z \sim 5$ these increase from 12+log(O/H) $\sim 7.3 \, (3.5\%~\zsun)$ for $M_* = 10^{7.5}\msun$ to 12+log(O/H) $\sim 8.0 \, (18\%~\zsun)$ for the most massive systems with $M_* \sim 10^{11}\msun$. As expected, the stellar mass range probed decreases with redshift such that by $z \sim 10$, the most massive galaxies with $M_* \sim 10^{9.4}\msun$ have a metallicity of 12+log(O/H) $\sim 7.7 \, (9\%~\zsun)$; the metallicity increases slightly (by about 1\%) at the low-mass end compared to that at $z \sim 5$. Further, at all redshifts, the metallicities show a $1-\sigma$ spread of about 0.15 dex around the median. 
All things considered, there is very little redshift evolution in the overall normalization of the MZR as shown in Fig. \ref{fig:mass_metallicity_hist_p}.

However, as also noted in Sec. \ref{sec:sfr} the underlying distribution shows metallicity values that span over 1.5 dex at the low-mass end. For example, $M_* \sim 10^{7.5}\msun$ galaxies show 12+log(O/H) values between 6.75-8.4 at $z \sim 5$ as a result of the cumulative effects of both SN and reionization feedback on their assembly histories. This range narrows to 12+log(O/H) $\sim 6.5-7.9$ at $z \sim 10$ given the fewer progenitor generations and the impact of reionization feedback on fewer galaxies. This metallicity range also narrows with an increase in stellar mass as their progenitors hold on to, and propagate, a larger fraction of their gas and metal mass to successive generations.

We also study the effects of maximal reionization feedback on the MZR in Fig. \ref{fig:mass_metallicity_hist_p}: as seen, the median metallicity values (and their scatter) at a given stellar mass do not show any sensible difference between the {\it Photoionization} and {\it Jeans mass} models at any redshift. This is due to the fact that the impact of the {\it Jeans mass model} is the most pronounced on the lowest mass ($M_* \lsim 10^{6.5}\msun$) galaxies that are hosted in $M_h \lsim 10^{9.2}\msun$ halos.

As has been noted in previous works \citep[e.g.][]{mannucci2010, lara-lopez2010}, the MZR is a two-dimensional projection of the 3D fundamental metallicity relation (FMR) that links the stellar mass, instantaneous SFR and the gas-phase metallicity. 
Furthermore, depending on the degree of curvature or saturation at the metal-rich end, this can be reduced to a fundamental plane of metallicity as shown by multiple works
\citep[e.g.][]{hunt2012, hunt2016, tortora2022}. We have performed a multiple linear regression, finding that our data lie on a 4-dimensional parameter space  - composed of the gas-phase metallicity, SFR, stellar mass and redshift - that can be expressed by a high-$z$ Fundamental Plane of Metallicity (HFPZ). We compute this FPZ by taking the central value for the stellar mass value in a bin and the median SFR in that particular bin. For the \textit{Photoionization} model the HFPZ is given by:
\begin{equation}
\begin{split}
12 + {\rm log(O/H)} = &- 0.294 \ {\rm log} \left( \frac{\rm SFR}{\Msun \rm{yr}^{-1}} \right) + 0.581 \ {\rm log} \left( \frac{M_*}{\Msun} \right)\\
                      &+ 2.272 + 0.061 \ z,
\end{split}
\label{eq:hfpz}
\end{equation}
while for the \textit{Jeans Mass} model the relation changes slightly to:
\begin{equation}
\begin{split}
12 + {\rm log(O/H)} = &- 0.342 \ {\rm log} \left( \frac{\rm SFR}{\Msun \rm{yr}^{-1}} \right) + 0.586 \ {\rm log} \left( \frac{M_*}{\Msun} \right)\\
                      &+ 2.216 + 0.061 \ z.
\end{split}
\end{equation}
As seen from the above equations, the normalisation of the HFPZ shows an extremely weak redshift evolution - this is discussed in more detail in Sec. \ref{sec:zev} that follows.

\subsubsection{Comparison against observational data}
We now compare our theoretical FPZ to those inferred from observations. The first dataset considered is the MEGA$^2$ sample that is introduced in 
\citet{tortora2022}.
This consists of $\sim 2100$ star forming galaxies with stellar
masses in the range $\sim 10^{6}-10^{11}\msun$, SFRs between $10^{-4}-10^3 \msun~{\rm yr^{-1}}$ and nebular oxygen abundance measurements ranging between 12+log(O/H)$\sim 7-9$ up to redshifts $z \sim 3.7$. Collected from 26 subsamples in the literature, this sample supersedes the previous MEGA sample \citep{hunt2016}. When necessary, stellar masses and SFRs are converted to a common \citet{chabrier2003} IMF according to \citet{speagle14}; for most of the samples SFRs are determined by using H$\alpha$, suitably corrected for extinction. For metallicities, when direct method estimates are not available, strong-line metallicity linear calibration for NII from \citet{pettini2004} have been used, converted from the original calibration if necessary following \cite{kewley2008}. 

The shape, or degree of curvature, of the MZR depends on many factors, including the metallicity calibration, the selection function, and the stellar mass distribution. Strictly speaking, a principal component analysis (PCA), is only applicable to samples without significant curvature, or saturation, at the high mass end. This is the case for typical high-redshift galaxy samples including  MEGA$^2$ \citep{tortora2022}.
Thus, similarly to \citet{hunt2012,hunt2016}, 
\citet{tortora2022} have performed a PCA on the MEGA$^2$ sample and confirmed that, 
independently of the O/H calibration and the redshift range, most of the variance is contained in the first two eigenvectors ($M_*$ and SFR) which comprise $98$\% of the total variance. Therefore, the metallicity can be predicted for any combination of these two parameters (i.e., $M_*$ and SFR). The best estimate for O/H as a function of $M_*$ and SFR from the MEGA$^2$ PCA, with an accuracy of $\pm 0.18$ dex, recalibrated to a Salpeter $0.1-100\msun$ IMF, is given by:
\begin{equation}
\begin{split}
12 + {\rm log(O/H)} = &- 0.10 \ {\rm log} \left( \frac{\rm SFR}{\Msun \rm{yr}^{-1}} \right) + 0.32 \ {\rm log} \left( \frac{M_*}{\Msun} \right)\\
                      &+5.35.
\end{split}
\end{equation}

Although the MEGA$^2$ O/H redshift dependence of the FPZ is not formally significant, there are indications that metallicity does decrease with redshift beyond what is predicted by the FPZ.
Analyzing the residuals of the FPZ as a function of redshift and recalibrating to a Salpeter $0.1-100\msun$ IMF, they obtain:
\begin{equation}
\begin{split}
12 + {\rm log(O/H)} = &- 0.10 \ {\rm log} \left( \frac{\rm SFR}{\Msun \rm{yr}^{-1}} \right) + 0.32 \ {\rm log} \left( \frac{M_*}{\Msun} \right)\\
                      &+ 5.45 - 0.039 \ z.
                      \label{eq:tor0}
\end{split}
\end{equation}

The second data set considered is that from Sanders et al. (2020) who fit the FMR using a set of high-$z$ O/H measurements between $z \sim 1.5-3.5$. However, unlike 
\citet{tortora2022}, they have imposed a different metallicity calibration for $z \sim 0$ and $z \ga 1$. Thus, in general, they find
a shallower evolution of the FPZ with redshift. The best-fit FMR found by \citet{sanders2020} for $M_* \sim 10^{9-10.5}\msun$ galaxies, recalibrated to a $0.1-100\msun$ Salpeter IMF (Sanders, private comm.) has the following functional form\footnote{Note that in \citet{sanders2020}, the $z \sim 2.3$ and $z \sim 3.3$ stacks fall directly on the $z \sim 0$ FMR despite the high-redshift stacks not being included in the fitting.}:
\begin{equation}
\begin{split}
12 + {\rm log(O/H)} = & 8.80 + 0.188 \ y - 0.220 \ y^2 - 0.0531 \ y^3,\\
{\rm where} \,\, y =                   & {\rm log} \left( \frac{M_*}{\Msun} \right) - 0.60 \ {\rm log} \left( \frac{\rm SFR}{\Msun \rm{yr}^{-1}} \right) - 10.1.
\end{split}
\end{equation}

In addition to these extrapolated fits, we also compare to the mass-metallicity relations observed for specific galaxies at $z \sim 7-9$ by \citet{jones2020}.

We now compare our FPZ with these observational relations in Fig. \ref{fig:mass_metallicity_hist_p}; 
these are 2D projection of a 4D relation composed of $M_*$, SFR, 12+log(O/H) and $z$. We 
caution that these comparisons are mostly intended as a sanity check of the overall normalization, given that they are extrapolated to such high-$z$ whilst being constrained only at $z \lsim 3.5$. 
At all redshifts, the extrapolated z-dependent FPZ from 
\citet{tortora2022} has a shallower slope and larger amplitudes compared to the theoretical model. 
In addition, as redshift decreases, the discrepancy between the observational and theoretical FPZ grows: at $z \sim 10$, the amplitude of the theoretical FPZ is lower than the observed $z$-dependent FPZ by only 0.15 dex, but this increases to 0.5 dex by $z \sim 0.5$ (at $M_* \sim 10^7\msun$). 

The $z$-independent ($z=0$) FPZ from the same authors essentially shows the same trends albeit the normalisation of the theoretical FPZ is under-estimated by much larger values of 0.5 and 0.7 dex at $z \sim 10$ and 5, respectively. The slope of the \citep{sanders2020} relation is in quite good agreement with the results from 
\citet{tortora2022},
although they find an even higher amplitude (by about 0.6 dex) compared to the $z$-independent relation shown in Eqn. \ref{eq:tor0}. Finally, within error bars, the results obtained for (the 4) individual galaxies from \citet{jones2020} are in agreement with ours at $z \sim 7$ and $8$; their $z \sim 9$ galaxy has a higher metallicity than our predicted upper limit by about 0.3 dex. 

\begin{figure*}
	\centering
	\includegraphics[width=0.9\linewidth]{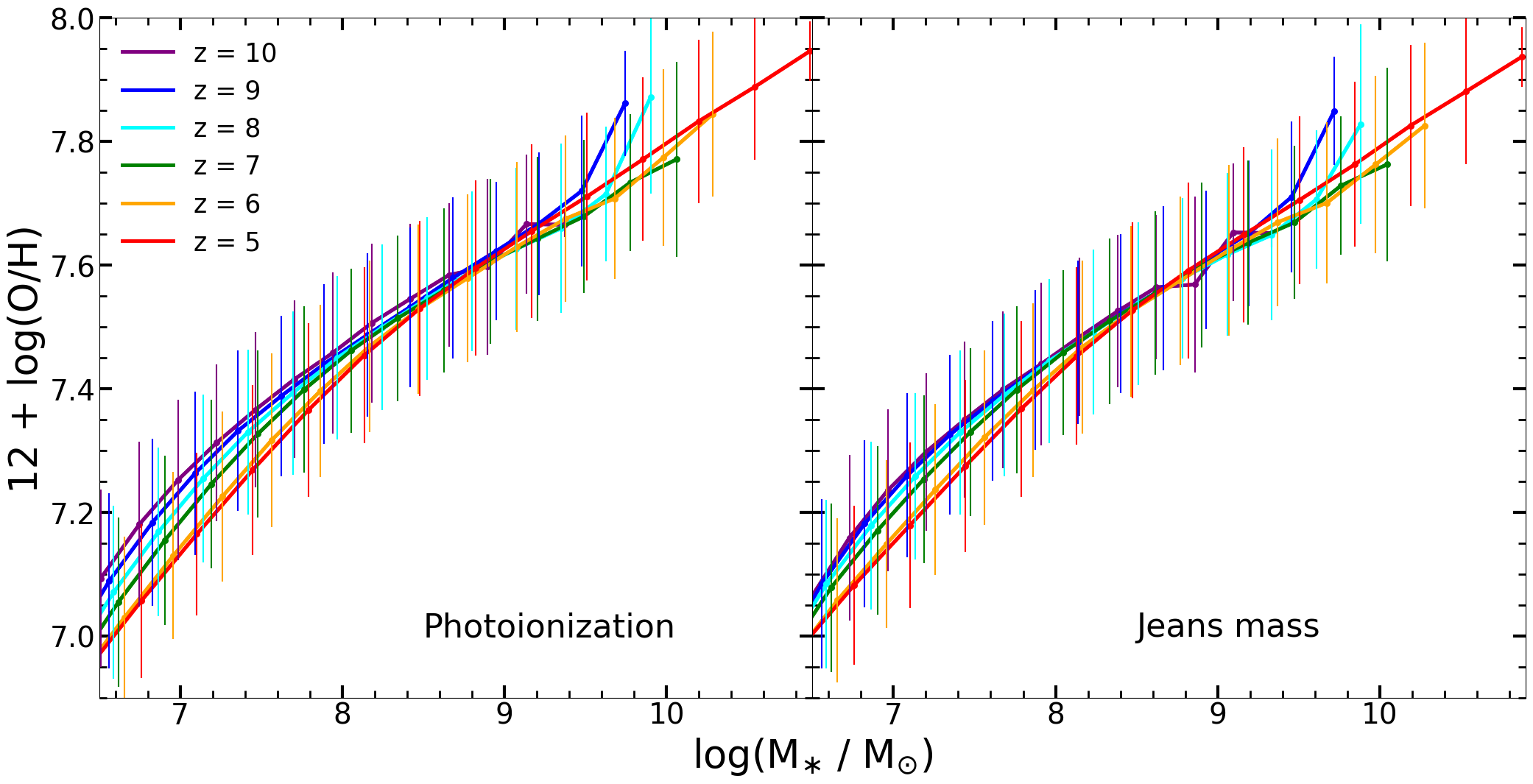}
	\caption{The redshift evolution of the mass-metallicity relation at $z \sim 5-10$. We show the median gas-phase metallicity as a function of the stellar mass for the redshifts marked, with the error bars showing the $1-\sigma$ dispersion. The {\it left} and {\it right} panels show results for the \textit{Photoionization} and \textit{Jeans mass} models, respectively. }
	\label{fig:mass_metallicity_z}
\end{figure*}

Reconciling the theoretical FPZ with these extrapolated observations would require theoretically-modelled galaxies to become progressively metal enriched with decreasing redshift (whilst maintaining given gas and stellar mass). This would require: (i) fewer metals being ejected per unit SFR i.e. a lower mass loading for metals at all masses. In order to maintain the gas mass, this might require ejected gas to be {\it preferentially metal poor} as compared to the ISM metallicity; (ii) a larger mass of metals to be gained through IGM accretion. This is along the lines of the ``galactic fountain" model where metal enriched gas ejected in previous time-steps in re-accreted onto the galaxy at a later stage \citep[e.g.][]{dave2011,muratov2017,oppenheimer2018}; or (iii) a combination of these two effects. 

\subsubsection{Comparison against theoretical models}
We now briefly compare our mass-metallicity relation at $z \sim 6$ to those found by a number of other theoretical models including \textsc{fire} \citep{ma2016},  \textsc{illustris tng} \citep{torrey2019} and \textsc{FirstLight} \citep{langan2020}, the results of which are shown in Fig. \ref{fig:comparison}. \textsc{fire} and \textsc{FirstLight} are both zoom-in simulations that track the mass-metallicity relation for $M_* \sim 10^{3-9.1}\msun$ and $M_* \sim 10^{6.25-9.25} \msun$ galaxies at $z \sim 6$, respectively. They both model SN yields \citep[from][]{woosley1995} alone\footnote{Note these are a factor 0.4 dex lower than the more recent yields from e.g. \citet{nomoto2013}.} and employ a solar metallicity normalisation (in units of 12+{\rm log} O/H) of 8.9. However, they differ in a number of key quantities such as the threshold gas density for star formation ($>1 {\rm cm^{-3}}$ and $10-100{\rm cm^{-3}}$ for \textsc{fire} and \textsc{FirstLight}, respectively), their implementation of feedback that severely impacts the gas and metal contents of galaxies and the IMF; this is a Kroupa IMF \citep{kroupa1993} in \textsc{fire} and Salpeter for \textsc{FirstLight}. Despite differences in both the physics implemented and the numerical methodology used, it is heartening to note that the results from these two simulations bracket ours: the results from \textsc{fire} and \textsc{FirstLight} are in good agreement with ours for $M_* \gsim 10^{8-9}\msun$ and $M_* \lsim 10^8\msun$ galaxies at $z \sim 6$, respectively. Finally, \textsc{illustris tng} has simulated a cosmological 100 $h^{-1}$ Mpc box including the key processes of gas cooling, star formation (at gas densities $>0.13 {\rm cm^{-3}}$), gas heating and feedback. This simulation uses the \citet{chabrier2003} IMF and includes metal production from SN as well as AGBs; although we caution that they use a solar normalisation (in units of 12+{\rm log} O/H) of 8.6. These last set of simulations are closest to ours in spirit in terms of the range of galaxy masses modelled and the metal-enrichment implemented. At $z \sim 6$, in the range of overlap with our results ($M_* \sim 10^{8-9}\msun$), these simulations predict metallicity values that are consistently higher than ours by about 0.5 dex. This indicates that: (i) we are either ejecting too many metals from the galaxy, which is reasonable given our ``maximally SN feedback limited" model; or (ii) that we gain too few metals through smooth accretion; this again is a plausible solution since we assume all gas accreted from the IGM to be metal-free. Indeed, as pointed out by \citet{ma2016}, the \textsc{fire} simulations clearly show that that outflowing metals can easily be retained between $0.25-1$ virial radii.

\begin{figure}
	\centering
	\includegraphics[width=1.0\linewidth]{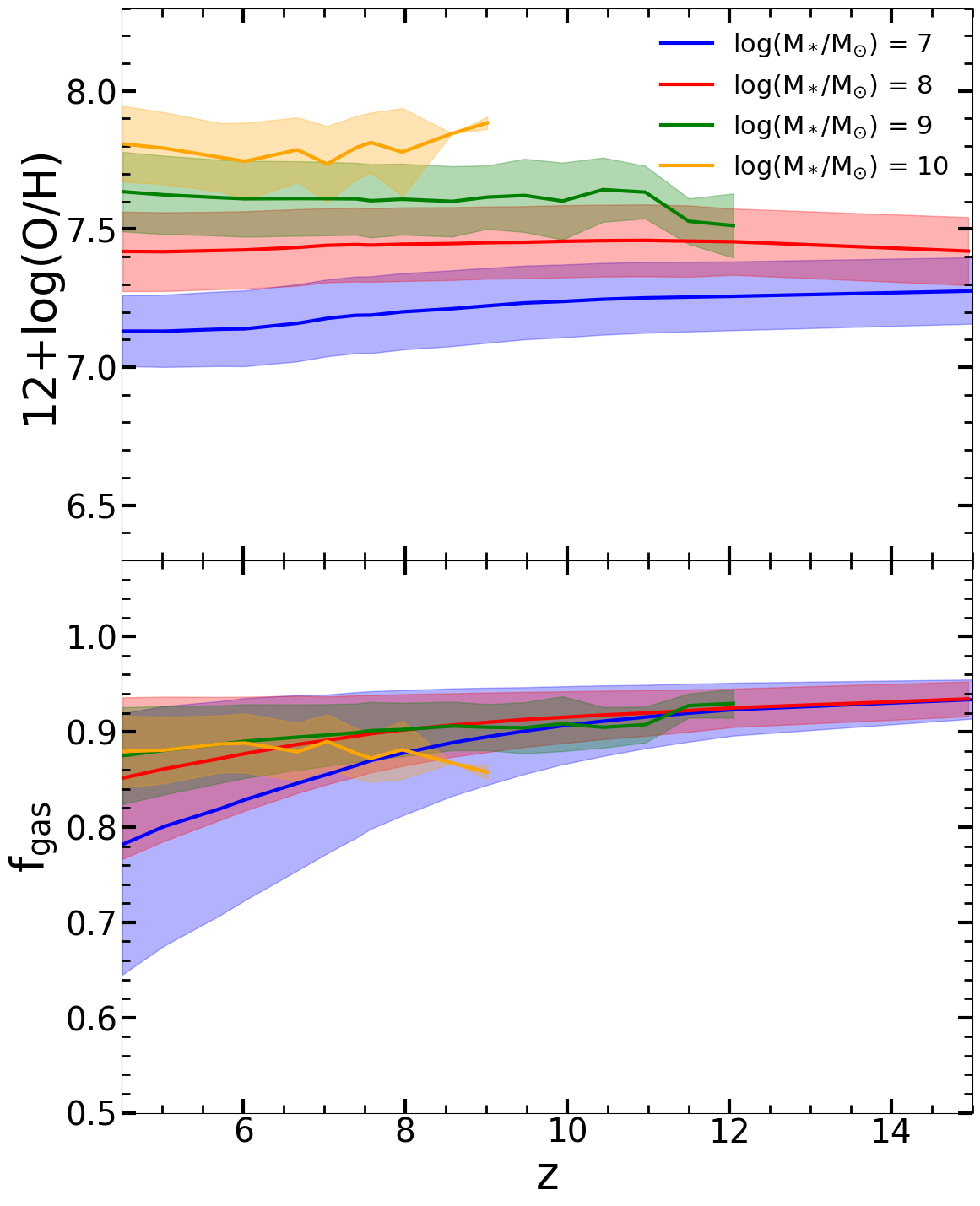}
	\caption{Redshift evolution of the average gas-phase metallicity (\textit{upper panel}) and average gas fraction $f_{\rm gas} = {\rm M}_{\rm gas} / ({\rm M}_{\rm gas} + {\rm M}_{\rm star})$ (\textit{lower panel}) for the stellar bins marked in the top panel.}
	\label{fig:mean_met}
\end{figure}
\subsection{The redshift evolution of the mass-metallicity relation}
\label{sec:zev}

We now discuss the redshift evolution of the mass metallicity relation using Fig. \ref{fig:mass_metallicity_z} that shows the median metallicity in each stellar mass bin (i.e. the median points as reported in Fig. \ref{fig:mass_metallicity_hist_p}) between $z \sim 5-10$.  As seen from this figure, a mass-metallicity relation is already in place at $z = 10$ and persists, albeit with a weak evolution, down to redshift $z = 5$. At a given stellar mass, the median metallicity decreases by about $0.15$ dex for $M_* \lsim 10^7 \msun$ galaxies. This difference decreases with increasing stellar mass such that the metallicity values converge by $M_* \sim 10^9 \msun$ at $z \sim 5-10$. This behaviour holds true for both the {\it Photoionization} and {\it Jeans mass} models.

The decrease in median metallicity with decreasing redshift for low-mass galaxies can be explained as follows: in our model, the stellar masses are directly linked to the underlying potential. At a given halo mass, galaxies can form stars with a slightly higher efficiency with increasing redshift as a result of their deeper potentials \citep[see Fig. 1][]{dayal2013}. This results in galaxies of a given stellar mass being hosted in halos that are 1.5 times more massive at $z \sim 5$ as compared to $z \sim 10$. Galaxies of similar masses at lower redshifts assemble from larger generations of feedback-limited progenitors that bring in little/no gas and metal content whose loss can not be fully compensated by accretion/mergers/self-enrichment; this decreases the metallicity with decreasing redshift. The decreasing impact of feedback with increasing halo mass results in the metallicity converging for larger mass galaxies with $M_* \sim 10^9 \msun$ (hosted in halos of mass $\sim 10^{11}\msun$). The {\it Jeans mass} model shows a smaller difference in metallicity at $z \sim 5-10$. This is probably driven by the fact that galaxies are more UV feedback suppressed in terms of their gas mass (and hence SFR and metal enrichment) in this model. This results in galaxies of a given stellar mass being hosted in halos that are 2.5 times more massive at $z \sim 5$ as compared to $z \sim 10$. This leads to slightly higher SFRs at $z \sim 5$ which are accompanied by more metal production, a larger fraction of which can be retained in the halo. This naturally decreases the difference between the median metallicity for a given stellar mass as a function of redshift. However, we caution the reader that considering the $1-\sigma$ errors, our model is consistent with {\it no redshift evolution of the mass-metallicity relation}.

A number of other works \citep{ma2016, torrey2019, langan2020} have used the gas fraction $f_{\rm gas} = {\rm M}_{\rm gas} / ({\rm M}_{\rm gas} + {\rm M}_*)$ to explain the redshift-evolution of the mass metallicity relation. 
Driven by the evolution of the gas fraction, the average relation predicted by \textsc{fire} \citep{ma2016} predicts the mass-metallicity relation to decrease by $\sim 0.1$ dex between $z \sim 5-10$ while \textsc{FirstLight} \citep{langan2020} predict the relation to be redshift-independent; these are in accord with our results that also show no redshift evolution of the mass-metallicity relation within $1-\sigma$ errors. On the other hand, \textsc{illustris tng} \citep{torrey2019} find that this relation decreases by about $0.3$ dex between the same redshifts, although their results are limited to $M_* \lsim 10^{9.5}\msun$ galaxies. 

For a comparison, we calculate the gas fraction and metallicity as a function of redshift for different stellar mass bins, as shown in Fig. \ref{fig:mean_met}. As seen from this plot, and discussed above, the average metallicity decreases by about 0.1 dex between $z \sim 15$ and $5$ for low-mass galaxies with $M_* \sim 10^7 \msun$. However, for larger masses (i.e. $M_* \gsim 10^8 \msun$), the metallicity evolution is effectively independent of redshift. This behaviour is also reflected in the gas fraction plot (lower panel of the same figure) where, for a given stellar mass, the gas fraction decreases with decreasing redshift: for $M_* \sim 10^7 \msun$ galaxies, $f_{\rm gas}$ decreases by about 0.1 dex between $z \sim 10$ and $5$. This essentially reflects the impact of the merger of successively feedback-limited systems on the final host halos. As might be expected, the gas fractions flatten out as a function of redshift for more massive systems, with $M_* \gsim 10^8\msun$ which is reflected in the redshift-independence of their metallicity.

\subsection{The metal enrichment of the IGM in the first billion years}
\label{sec:igm_met}
\begin{figure}
	\centering
	\includegraphics[width=1.0\linewidth]{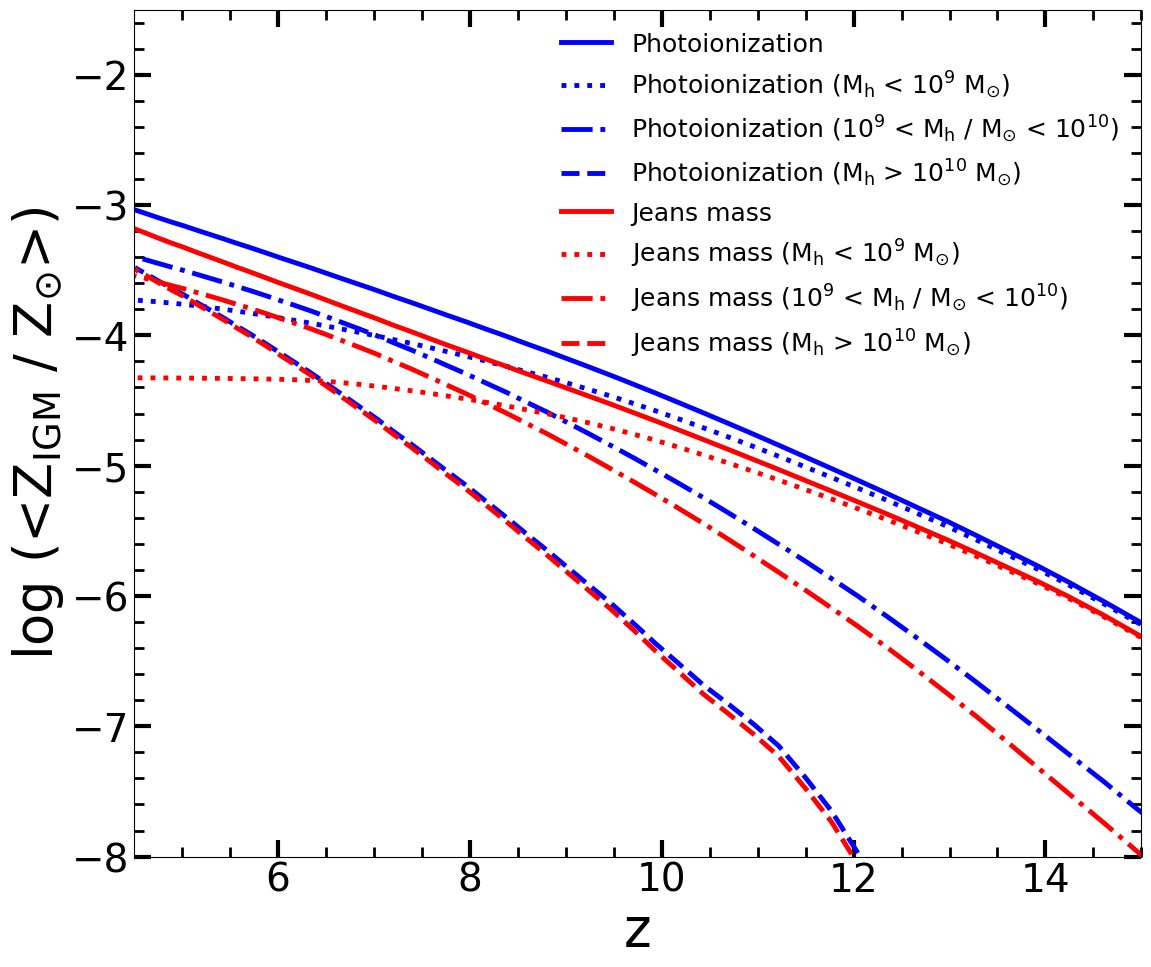}
	\caption{Redshift evolution of the average IGM metallicity (in solar units) in the entire simulation box. The solid blue and red lines show results for all galaxies in the \textit{Photoionization} and \textit{Jeans mass} models, respectively. As marked, these have been deconstructed into the contribution from low-mass halos ($M_h \lsim 10^9\msun$; dotted lines), intermediate-mass halos ($M_h \sim 10^{9-10}\msun$; dot-dashed lines) and high-mass halos ($M_h \gsim 10^{10}\msun$; dashed lines).}
	\label{fig:Zigm}
\end{figure}

Finally, we discuss the metal enrichment of the IGM at $z\gsim 4.5$ for both the \textit{Photoionization} and \textit{Jeans mass} as shown in Fig. \ref{fig:Zigm}. 
This is the {\it average} IGM metallicity obtained by diving the total metal mass in the IGM (i.e. outside the virial radii of all halos) by the total gas mass (see Eqn. \ref{eqn_zigm}). We caution that the IGM metallicity values used here must be treated as a lower limit since the ejected metals are assumed to be homogeneously dispersed into the entire simulation box when calculating $Z_{\rm IGM}$. In principle, metals might be expected to be preferentially clustered around the galaxies that eject them. These metals can then be accreted onto their host galaxies in a future time-step (the ``galactic fountain'' model) resulting in the accretion of metal-enriched gas \citep{dave2011,muratov2017,oppenheimer2018}.

Starting with the {\it Photoionization} model, the IGM metallicity has a value of about $10^{-6.2}\zsun$ at $z \sim 15$. The formation of an increasing number of low to intermediate mass galaxies, that form and eject metals into the IGM, with cosmic time naturally results in a corresponding increase in the metallicity. Indeed, by $z \sim 4.5$, $Z_{\rm IGM}$ increases by about 3 orders of magnitude to $\sim 10^{-3}\zsun$. We then deconstruct this into the contribution from low-mass ($M_h < 10^9\ \Msun$), intermediate ($10^9 < M_h /\Msun < 10^{10}$) and high-mass ($M_h > 10^{10}\ \Msun$) halos as shown in the same figure. The IGM enrichment is dominated by low mass halos at $z \gtrsim 8$ both because of their higher number densities and the fact that these objects are SN feedback dominated and hence able to expel essentially all of their metal content into the IGM. Indeed, their contribution to the IGM metallicity is of the order of $\sim 90\%$ at $z \sim 12$. Once reionization in underway, their contribution starts decreasing as the gas content of a larger fraction of such sources is suppressed due to radiative feedback, leading to a corresponding decrease in the star formation rate and metal production. Indeed, their contribution to the total IGM metal budget is of the order of 50\% at $z \sim 7$ - this also corresponds to the redshift at which roughly half of the IGM is ionized \citep[see Fig. 9;][]{astraeus3}; thereafter their contribution drops even faster due to radiative feedback such that these galaxies contribute $\sim 15\%$ to the IGM metal budget by $z \sim 4.5$. Intermediate-mass halos are still able to expel metal-enriched winds, although they can retain more of their metal mass which results in a $\sim10$\% contribution at $z \sim 12$. Since these numerous systems are less affected by radiative feedback, their contribution to the IGM metallicity starts being significant at $z \lesssim 7$ - by $z \sim 4.5$, such systems contribute $\sim 56\%$ to the IGM metallicity. Finally, as a result of their low number densities and larger potentials, high-mass galaxies have essentially no contribution at $z \sim 12$. As an increasing number of such systems assemble with decreasing redshift, their contribution to the metal budget shows a steep increase such that they are responsible for $\sim 28\%$ of the metal budget by $z \sim 4.5$. 

Qualitatively, the results from the {\it Jeans mass} model are very similar. However, this model has a  much stronger (instantaneous) feedback effect that suppresses the gas fraction (and hence star formation) in low-intermediate mass galaxies to a larger extent \citep{astraeus1}. As a consequence, such galaxies are both able to produce and eject a smaller amount of gas and metals into the IGM as compared to the {\it Photoionization} model. This leads to a slightly lower IGM metallicity (by about 0.2 dex) as compared to the {\it Photoionizaton} model at effectively all redshifts. As might be expected, low-mass galaxies still dominate the IGM metallicity contribution at early times. However as a result of the stronger radiative feedback, their contribution drops to the 50\% level as early as $z \sim 10$ in this model. Thereafter, the contribution of such galaxies effectively flattens, such that they contribute $\sim 10\%$ to the IGM metallicity by $z \sim 4.5$. The contribution of intermediate-mass galaxies (that are much less affected by radiative feedback), is about 0.2 dex lower at all redshifts in this model as compared to the {\it Photoionization} model. Such galaxies contribute about $40\%$ ($45\%$) to the IGM metallicity at $z \sim 7$ (4.5). Finally, high-mass galaxies, that are essentially unaffected by radiative feedback, have the same absolute contribution in both radiative feedback models. In the {\it Jeans mass} model, such galaxies contribute about $10\%$ ($45\%$) to the IGM metallicity at $z \sim 7$ (4.5).
\section{Conclusions and discussion}
\label{sec:conc}
In this work, we investigate the chemical enrichment of early ($z \gsim 5$) galaxies and the emergence of metallicity scaling relations using the \textlcsc{Astraeus} framework that self-consistently couples galaxy formation and reionization using an N-body simulation ({\sc very small multidark planck; vsmdpl}), a semi-analytic model for galaxy formation (an extended version of {\sc Delphi}) and a semi-numerical radiative transfer code for reionization ({\sc cifog}). The key strength of our model lies in: (i) the radiative feedback models explored  that range from a weak, time-delayed one (the {\it Photoionization} model) to a strong instantaneous reduction of gas in the galaxy (the {\it Jeans mass} model); and (ii) the fact that \textlcsc{Astraeus} is the only semi-analytic model to include the latest state-of-the-art yields from \citet{kobayashi2020} which reproduce the observations not just for oxygen but also for most of all stable elements (up to uranium) self-consistently. Our key findings for $z \gsim 4.5$ galaxies are:

\begin{itemize}
    \item The mass brought in by merging progenitors dominates the assembly of both the gas and metal contents at all redshifts, followed by smooth accretion from the IGM and SNII-driven ejection from the ISM. The gas returned by exploding stars is negligible, being about 2.5 orders of magnitude less than the merged gas mass.
    
    \item Mergers also dominate the metal assembly, followed by self-enrichment and ejection. Given that the IGM metallicity only reaches a value of about $Z_{\rm IGM} \sim 10^{-3}\Msun$ by $z \sim 4.5$, smooth accretion does not have any sensible contribution to the metal assembly.
    
    \item Irrespective of the stellar mass, the smooth-accretion of metal-poor gas from the IGM plays a key role in {\it diluting} the ISM metallicity which is effectively restored due to self-enrichment from star formation. As expected, given our assumption of gas and metals being perfectly mixed, ejection has no impact on the final metallicity. 
    
    \item For the {\it Photoionization model}, the average gas mass loading factor scales with the stellar mass, for $M_* \sim 10^{7-10.4}\msun$  ($M_* \sim 10^{7-9.5}\msun$) galaxies, at $z \sim 5$ (10) such that 
\be
\eta_g \approx 1.38 \bigg(\frac{M_*}{10^{10}\msun}\bigg)^{-0.43}.
\ee
    
     \item Interestingly, we find that the stellar mass-gas phase metallicity relation (MZR) was already in place as early as $z \sim 10$ and persists, effectively un-evolving, down to $z \sim 5$. 
     
     \item We also find a three-dimensional correlation between the metallicity, stellar mass and star formation rate (the fundamental metallicity relation; FMR) to persist at all $z \sim 10-5$. Essentially, for a given stellar mass, the metallicity decreases with an increase in the star formation rate. However, given that self-enrichment supersedes dilution at all the masses studied, we do not see any flattening of the mass-metallicity relation for high-masses, as has been observed at lower-redshifts ($z \sim 0$).
     
      \item We find our simulated galaxies to lie on a 4-dimensional relation [12+log(O/H), SFR, $z$, $M_*$] that can be expressed by a relation we named \quotes{high-z Fundamental Plane of Metallicity} (HFPZ). Performing a multiple linear regression, we find that the HFPZ is given by:
    \begin{equation}
    \begin{split}
    12 + {\rm log(O/H)} = &- 0.294 \ {\rm log} \left( \frac{\rm SFR}{\Msun \rm{yr}^{-1}} \right) + 0.581 \ {\rm log} \left( \frac{M_*}{\Msun} \right)\\
                          &+ 2.272 + 0.061 \ z
    \end{split}
    \end{equation}
    and
    \begin{equation}
    \begin{split}
    12 + {\rm log(O/H)} = &- 0.342 \ {\rm log} \left( \frac{\rm SFR}{\Msun \rm{yr}^{-1}} \right) + 0.586 \ {\rm log} \left( \frac{M_*}{\Msun} \right)\\
                          &+ 2.216 + 0.061 \ z
    \end{split}
    \end{equation}
    for the \textit{Photoionization} and \textit{Jeans mass} models, respectively.\\
     
     \item Interestingly, we find that both the {\it Photoionization} and {\it Jeans mass} models lead to very similar qualitative results for both the gas and metal mass assembly as well as the mass-metallicity relations explored here; the impact of the {\it Jeans mass model} is the most pronounced on the lowest mass ($M_* \lsim 10^{6.5}\msun$) galaxies that are hosted in $M_h \lsim 10^{9.2}\msun$ halos.    
    
    \item The average IGM metallicity increases from about $Z_{\rm IGM} \sim 10^{-6.2}$ to $10^{-3}\zsun$ between $z \sim 15$ to 4.5. Most of this enrichment is driven by low-mass galaxies (with halo mass $M_h \lsim 10^9\Msun$) at $z \sim 8$ in the {\it Photoionization} model. As such low-mass galaxies are progressively suppressed by both SNII and radiative feedback, the contribution of intermediate to high-mass halos increases at $z \lsim 8$ such that they contribute $\sim 56\%$ and $\sim 28\%$ to the IGM metal budget by $z \sim 4.5$. Given the much stronger effect of radiative feedback, the contribution of low-mass galaxies drops off faster in the {\it Jeans mass} model; this is compensated by a higher contribution from intermediate-high mass galaxies.
    
\end{itemize}

We end by noting that, encouragingly, our results for the mass-metallicity relation are within the limits predicted by a number of theoretical models (including \textsc{fire}, \textsc{FirstLight} and \textsc{illustris tng}). However, we increasingly under-predict the metallicity with decreasing redshift for a given stellar mass, when compared to observationally-extrapolated results from $z \sim 0-3.5$. This could be due to a number of simplifying assumptions made in our model such as: (i) all of the gas mass can form stars; (ii) smoothly accreted gas having an IGM metallicity value that is averaged over the entire box rather than accounting for the distribution of metals in the IGM; (iii) a perfect mixture of gas and metals being ejected; (iv) ignoring the presence of dust in high-z galaxies. The last point is particularly relevant in light of recent Atacama Large Millimetre Array (ALMA) observations that show significant dust masses attenuating the UV light from relatively normal high-redshift star forming galaxies \citep{bouwens2021, fudamoto2021}. Indeed, this might require re-calibrating our threshold star formation efficiencies, especially for high-mass galaxies, impacting their metal masses. Forthcoming observations with the JWST will truly be crucial in shedding light on the high-redshift mass-metallicity relation and its redshift evolution.
 
\section*{Acknowledgements}
GU, PD, AH, LL, GY, SG and LH acknowledge support from the European Research Council's starting grant ERC StG-717001 (\quotes{DELPHI}). PD acknowledges support from the NWO grant 016.VIDI.189.162 (\quotes{ODIN}) and the European Commission's and University of Groningen's CO-FUND Rosalind Franklin program. CK acknowledges funding from the UK Science and Technology Facility Council (STFC) through grant ST/R000905/1 \& ST/V000632/1. GY acknowledges financial support from MICIU/FEDER under project grant PGC2018-094975-C21. The authors are grateful to Ryan Sanders for providing observational MZR results recalibrated to our IMF. 
We thank Peter Behroozi for creating and providing the \textlcsc{Rockstar} merger trees of the VSMPL simulation. The authors wish to thank V. Springel for allowing us to use the L-Gadget2 code to run the different Multidark simulation boxes, including the VSMDPL used in this work. The VSMDPL simulation has been performed at LRZ Munich within the project pr87yi. The authors gratefully acknowledge the Gauss Centre for Supercomputing e.V. (www.gauss-centre.eu) for funding this project by providing computing time on the GCS Supercomputer SUPERMUC-NG at Leibniz Supercomputing Centre (www.lrz.de). The CosmoSim database (\url{www.cosmosim.org}) provides access to the simulation and the Rockstar data. The database is a service by the Leibniz Institute for Astrophysics Potsdam (AIP). 

\section*{Data Availability}
The \textlcsc{Astraeus} simulations and derived data in this research will be shared on reasonable request to the corresponding authors.

\bibliographystyle{mnras}
\bibliography{metallicity}

\bsp
\label{lastpage}

\appendix

\section{The relation between the stellar mass, gas-phase metallicity and star formation rate}
\label{app:fmr}
\begin{figure*}[h]
	\centering
	\includegraphics[width=0.9\linewidth]{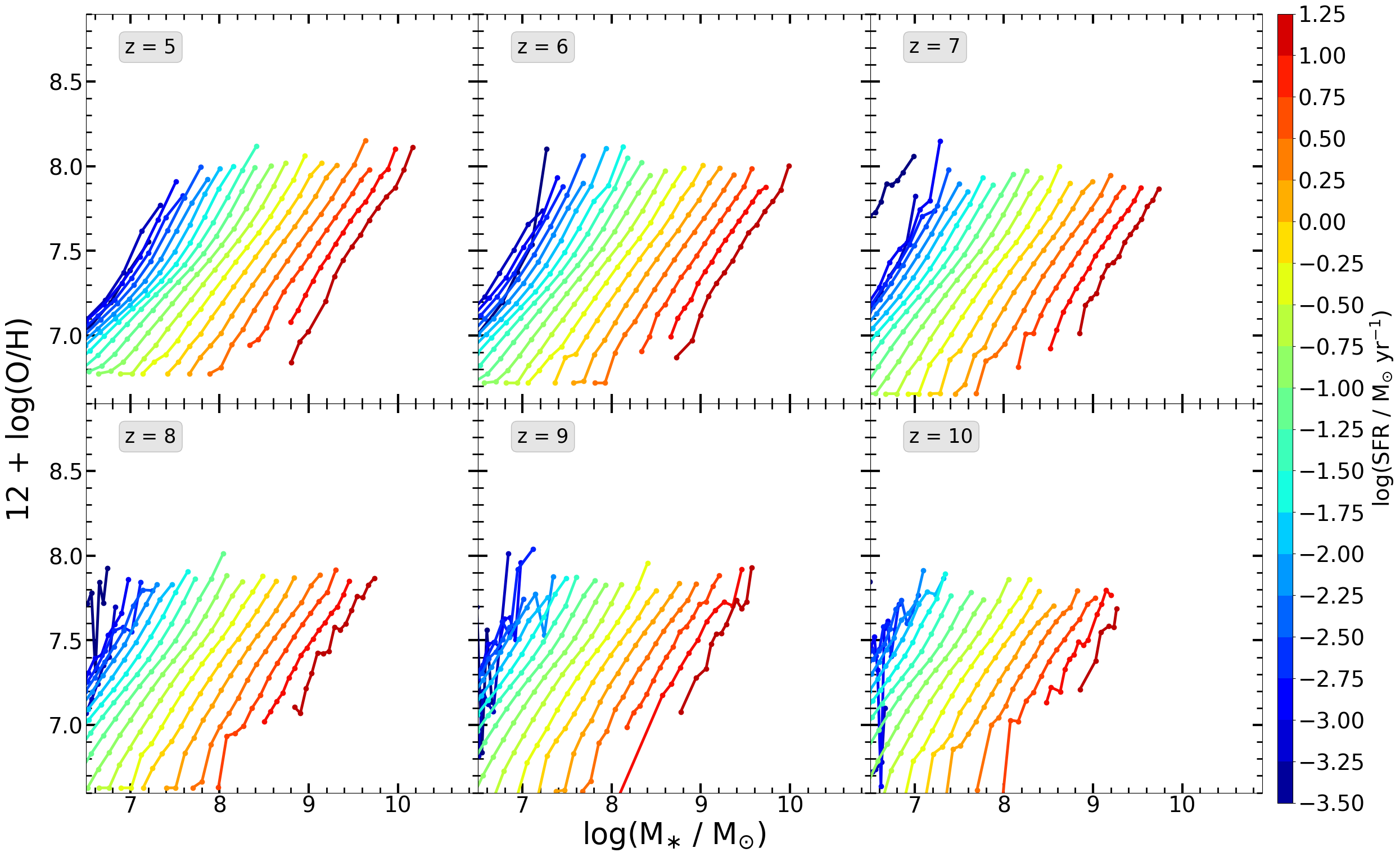}
	\caption{The (final) gas-phase metallicity, in units of $12+{\rm log~(O/H)}$, as a function of stellar mass for $z \sim 5-10$ as marked. In each panel, the results have been color-coded by the (log) star formation rate. For clarity, we only show results for the \textit{Photoionization} model given that the \textit{Jeans mass} model leads, visually, to very similar values.  }
	\label{fig:mass_metallicity_sfr_p}
\end{figure*}

In this section, we discuss the three-dimensional relation between the stellar mass, instantaneous star formation rate and the gas-phase metallicity - the three parameters comprising the FMR; qualitatively, these results are similar to those detailed in Sec. \ref{sec:sfr}. As seen in Fig. \ref{fig:mass_metallicity_sfr_p}, independent of redshift, at a given stellar mass the metallicity decreases with an increase in the star formation rate. For example, at $z \sim 5$,  for $M_* \sim 10^7 \Msun$ galaxies, the metallicity decreases from 12+log(O/H) $\sim 7.7$ to $\sim 6.8$ as the SFR increases from $10^{-2.5}$ to $10^{-0.75} \Msun ~ {\rm yr^{-1}}$. The same behaviour persists for galaxies as massive as $10^{9.5}\Msun$ where the metallicity value drops from 8 to 7.5 as the SFR increases from $10^{0.25}$ to $10 \Msun ~ {\rm yr^{-1}}$. While this trend is mostly driven by the {\it ejection} of metals at the low-mass end, {\it dilution} (required to sustain higher SFR) becomes the primary driver at the high-mass end. Interestingly, we do not see any flattening of the metallicity even for the highest mass galaxies. This is because self-enrichment still supersedes dilution. Although these same trends persist at all $z \sim 5-10$, the stellar mass and SFR ranges naturally decrease with increasing redshift. Finally, low-mass galaxies ($M_* \lsim 10^{7.6}\Msun$) with extremely high metallicities are again outliers where the only gas (metal) mass are those returned by exploding stars (self-enrichment in the last redshift step) as also noted in Sec. \ref{sec:sfr}.

\section{Comparison of the metallicity values derived from the HFPZ and {\sc Astraeus}}

\begin{figure*}
	\centering
	\includegraphics[width=0.9\linewidth]{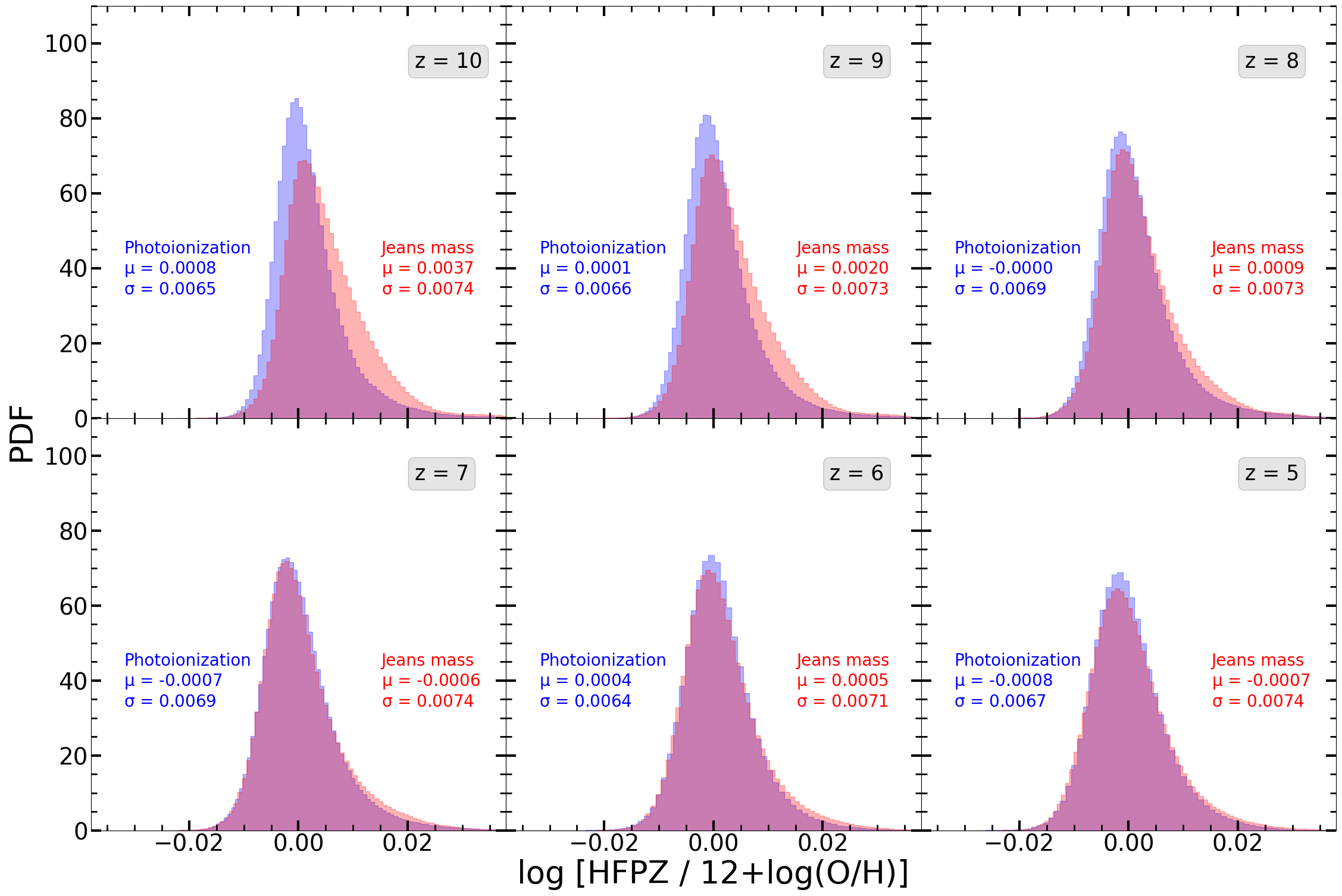}
	\caption{Histograms of ratios between the metallicity derived with the HFPZ described in Eqn. \ref{eq:hfpz} and the actual value of the metallicity obtained in our simulation at different redshift expressed in log (i.e., a value of 0 means that the actual value of the metallicity in our simulation and the one derived with the HFPZ are exactly the same). Blue and red shaded histograms show the results for the \textit{Photoionization} and \textit{Jeans mass} models, respectively along with their mean ($\mu$) and standard deviation ($\sigma$). As reference, a value of 0.03 implies a difference between the HFPZ and the 12+log(O/H) value of $\sim$ 7\%. In blue and red we also report the mean ($\mu$) and standard deviation ($\sigma$) for the distributions obtained for the \textit{Photoionization} and \textit{Jeans mass} model, respectively. }
	\label{fig:diffs}
\end{figure*}

In Fig. \ref{fig:diffs}, we show the histograms of the log of the ratios between our inferred HFPZ and the metallicity at different redshifts (a value of 0 means that the actual value of the metallicity and the one derived with our HFPZ are exactly the same)  in our simulation: the residuals are small across the redshift range $z = 5-10$ and almost the whole sample of our simulated galaxies have a metallicity that differs from the value inferred from the HFPZ less that $\sim$ 7\%.

\end{document}